\documentclass[onecolumn]{aastex62}
\usepackage{graphicx}
\usepackage{amssymb}                    
\usepackage{ulem}
\usepackage{float}
\usepackage{rotating}
\usepackage{enumitem}
\usepackage{makecell}
\usepackage{appendix}
\usepackage{fixmath}

\newcommand{\arc}{\ensuremath{^{\prime\prime}}}
\newcommand{\none}{\ensuremath{^{-1}}}

\usepackage{times} 
\usepackage{amsmath}
\received{March 18, 2019}
\submitjournal{ApJ}

\shortauthors{Winebarger et al.}
\shorttitle{UNFOLDING IMAGING SPECTROMETER DATA}

\begin{document}


\title{UNFOLDING OVERLAPPED SLITLESS IMAGING SPECTROMETER DATA FOR EXTENDED SOURCES}

\correspondingauthor{Amy Winebarger}
\email{amy.winebarger@nasa.gov}

\author{Amy R.\ Winebarger}
\affil{NASA Marshall Space Flight Center, ST13, Huntsville, AL 35812}

\author{Mark Weber}
\affil{Smithsonian Astrophysical Observatory, 60 Garden Street, Cambridge, MA 02138}

\author{Christian Bethge}
\affil{NASA Marshall Space Flight Center, ST13, Huntsville, AL 35812}
\affil{Universities Space Research Association, 320 Sparkman Drive, Huntsville, AL 35806, USA}

\author{Cooper Downs}
\affil{Predictive Science Inc., 9990 Mesa Rim Road, Suite 170, San Diego, CA 92121}

\author{Leon Golub}
\affil{Smithsonian Astrophysical Observatory, 60 Garden Street, Cambridge, MA 02138}

\author{Edward DeLuca}
\affil{Smithsonian Astrophysical Observatory, 60 Garden Street, Cambridge, MA 02138}

\author{Sabrina Savage}
\affil{NASA Marshall Space Flight Center, ST13, Huntsville, AL 35812}

\author{Giulio del Zanna}
\affil{DAMTP, Center for Mathematical Sciences, University of Cambridge, Wilberforce Road, Cambridge, CB3 0WA, UK}
\author{Jenna Samra}
\affil{Smithsonian Astrophysical Observatory, 60 Garden Street, Cambridge, MA 02138}
\author{Chad Madsen}
\affil{Smithsonian Astrophysical Observatory, 60 Garden Street, Cambridge, MA 02138}
\author{Afra Ashraf}
\affil{Barnard College, 3009 Broadway, New York, NY 10027}
\author{Courtney Carter}
\affil{Grinnell College, 1115 8th Ave, Grinnell, IA 50112}

\begin{abstract}
Slitless spectrometers can provide simultaneous imaging and spectral data over an extended field of view, thereby allowing rapid data acquisition for extended sources. In some instances, when the object is greatly extended or the spectral dispersion is too small, there may be locations in the focal plane where emission lines at different wavelengths contribute. It is then desirable to unfold the overlapped regions in order to isolate the contributions from the individual wavelengths. In this paper, we describe a method for such an unfolding, using an inversion technique developed for an extreme ultraviolet imaging spectrometer and coronagraph named the COronal Spectroscopic Imager in the EUV (COSIE).  The COSIE spectrometer wavelength range (18.6 - 20.5 nm) contains a number of strong coronal emission lines and several density sensitive lines.  We focus on optimizing the unfolding process to retrieve emission measure maps at constant temperature, maps of spectrally pure intensity in the Fe XII and Fe XIII lines and density maps based on both Fe XII and Fe XIII diagnostics.    
\end{abstract}
\keywords{Sun:corona}

\section{INTRODUCTION}

The ability to unfold spatial and spectral signals from an objective grating slitless spectroheliograph opens new windows into coronal spectroscopy.  Such instruments allow for simultaneous high spectral resolution observations across the entire field of view with a single exposure.  In the data from these instruments, sometimes called overlappograms, spatial and spectral information is convolved in the dispersive direction, making the interpretation of the data difficult.  

There is a long history of objective grating spectroscopy in solar physics. The Naval Research Laboratory (NRL) overlappograms from the NRL S-082A spectroheliograph on {\it Skylab} \citep{Tousey1973} opened up the coronal EUV window for spectroscopic analysis, and demonstrated the richness of the spectral signatures for different solar features. More recently the Res-K instrument of the Russian KRONOS-I mission \citep{Zhitnik1998} identified 51 lines in the spectral region of 18.0\,nm to 21.0\,nm. Res-K was optimized to provide high separation of the spectral lines at the expense of spatial resolution in the dispersive direction. Their spectrograph images had a factor of 10 difference in the spatial resolution in the dispersive direction compared with the cross-dispersive direction.  Kankelborg and collaborators \citep{Kankelborg01} designed, built, and flew the Multi-Order Solar EUV Spectrograph (MOSES), a novel objective grating telescope that simultaneously captures three spectral orders of the strong He II 30.4\,nm line. In their design the spatial information is captured in the zero order and the spectral information in the plus and minus one orders. Doppler shifts are easily detected in this design as the spectral displacements for the plus and minus one orders are in opposite directions.   Also, the Extreme-ultraviolet Imaging Spectrometer (EIS; \citealt{culhane2007}) is a scanning slit spectrometer that is currently operating on the {\it Hinode} satellite, but has two slot positions (40\arcsec and 266\arcsec wide) that can produce overlappogram data.

 Many methods of interpreting these data sets have been attempted.  For example,  \cite{fox10} performed a detailed analysis of a single explosive event observed in the MOSES data, fitting the intensity in the zero and first order and interpreting displacements in the first order as bulk flows and increased width in the first order as Doppler broadening.  \cite{courrier2018} uses Fourier Local Correlation Tracking to cross correlate the multiple image pairs observed by MOSES so that the displacements can be interpreted as velocity.  Additionally, \cite{harra2017} used intensities from an EUV imager to predict the spatial distribution of the emission in the EIS slot data so that additional displacement in the dispersive direction could be interpreted as a bulk flows or non-thermal velocities during the early stages of a flare eruption. 

A new instrument that makes use of objective grating spectroscopy is currently being proposed as a NASA Mission of Opportunity. It is called the ``COronal Spectroscopic Imager in the EUV: COSIE".  COSIE is a compact instrument that combines a wide field broadband EUV imager with an objective grating imaging spectrograph. COSIE is optimized for high throughput spatial imaging out to 3$R_\odot$ and high throughput spectroscopy.  COSIE data will capture global evolution and identify transient events that are difficult to capture with slit spectrographs.    The COSIE wavelength range is similar to Res-K (18.6-20.5\,nm), but for COSIE there is only a factor of three difference in spatial resolution between the spatial and dispersive directions (3.1"/pixel vs. 9.3"/pixel).

In this paper we apply unfolding methods developed for the MUlti-slit Solar Explorer (MUSE) mission \citep{cheung2018} to the 
COSIE data sets. The success of the approach for two substantially different instrumental setups is an indication of the robustness of the underlying method.
From the unfolded spectra we can measure at each pixel (with sufficient signal) the emission measure distribution (i.e. the electron temperature of the plasma), spectrally pure intensities in the strong lines in the wavelength range,  the electron density using line ratios from Fe XII \& XIII, and a spectrum along each line of sight.  We focus on the ability to unfold the data for a quiescent Sun that includes active regions and coronal holes.  We defer discussion of dynamic events and regions with strong velocities to a subsequent paper. In Section 2, we describe the COSIE instrument.  In Section 3, we describe the unfolding method.  We validate this  method using a 3D magnetohydrodynamic simulation in Section 4.  In Section 5, we demonstrate this method using a data set derived from EUV images.   In Section 6, we provide our conclusions and discuss how this technique can also be applied to other instruments and data sets.

\section{COSIE DESCRIPTION}

The COSIE optical system consists of a planar feed optic, spherical focus mirror, and 2k x 2k backside thinned CCD detector. Light at a steep incident angle is collected by the feed optic, directed to the focus mirror, and focused onto the detector through a hole in the feed optic. The conversion from coronagraph to spectrograph is performed by flipping the feed optic, which has a mirror on one side and a diffraction grating on the other. Figure 1 shows the light paths for the coronagraph and spectrograph channels.

\begin{figure}[ht]
\centering
{\includegraphics[width=.8\textwidth]{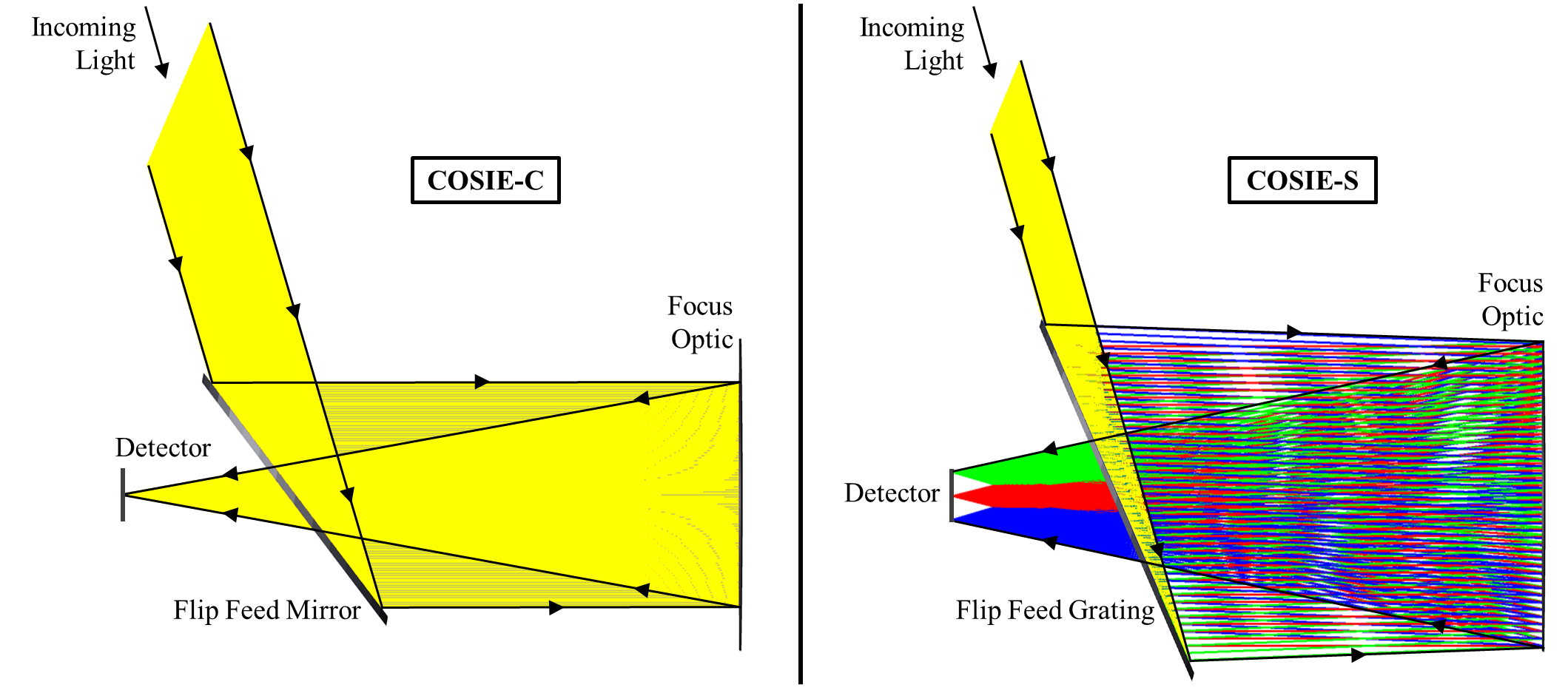}}    
\caption{Ray trace for the two COSIE Channels: a wide-field coronagraph (left) and slitless spectrograph (right)}
\label{fig:raytrace}
\end{figure}

The passband of the instrument, defined by the reflectances of the optical surfaces and the transmissions of the filters, is nominally 18.6-20.5 nm.  The off-disk coronagraph sensitivity is hundreds of times larger than the Geostationary Operational Environmental Satellite (GOES) Solar UltraViolet Imager (SUVI) and Solar Dynamics Observatory (SDO) Atmospheric Imaging Assembly (AIA) 19.3\,nm channels \citep{DelZanna18}.  To simultaneously image both the extended corona and the disk, a partially absorbing filter is placed in the focal plane to effectively reduce the on-disk emission in the EUV by $\approx$200 times, allowing for 1\,s  exposures that include both the on and off disk coronal features.   Taking into account the grating efficiency, the intensities of the individual spectral lines and the area of the grating, plus the lack of an absorbing disk in the overlappogram image, the expected exposure time for the spectrograph is also on the order of 1\,s. The effective areas of the coronagraph (including the absorbing filter) and spectrograph are given in Figure~\ref{fig:ea}.  {In the spectrograph, the spectral dispersion is $9.3 \times 10^{-4}$ nm/pixel and the spatial pixel size is 9.3\arcsec\ and 3.1\arcsec\ in the dispersive and cross-dispersive dimensions, respectively.  In the coronagraph, the spatial resolution is 3.1\arcsec\ with a field of view of $6.6 \times 6.6$ R$_\circ$.

\begin{figure}[ht]
\centering
{\includegraphics[width=.8\textwidth]{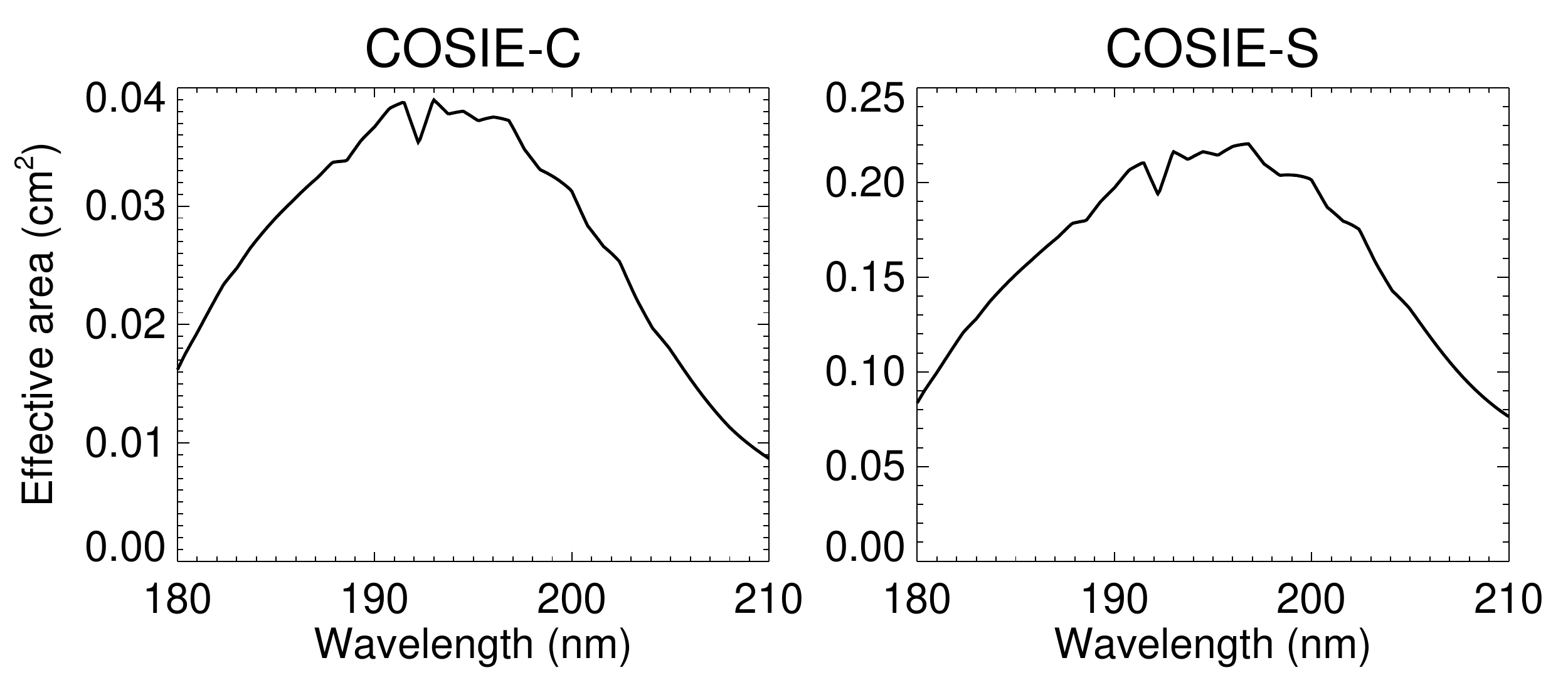}}    
\caption{The on-disk coronagraph effective area, with absorbing filter in place (left) and spectrograph effective area (right). }
\label{fig:ea}
\end{figure}

Examples of the COSIE-C and COSIE-S data products are shown in Figure~\ref{fig:COSIE-C-S}. The N-S axis of the Sun will be oriented along the dispersion direction in COSIE-S in order to reduce overlap from multiple active regions if there are several on the disk. The field of view extends out to 3$R_\odot$ and we expect there to be significant signal in the outer corona in the EUV (see \citealt{DelZanna18} for a detailed discussion.)  In this paper, we focus on the inner corona where we expect the spectroheliogram signal to be strong.

\begin{figure}[h!]
 \centering
  {\includegraphics[width=.9\textwidth]{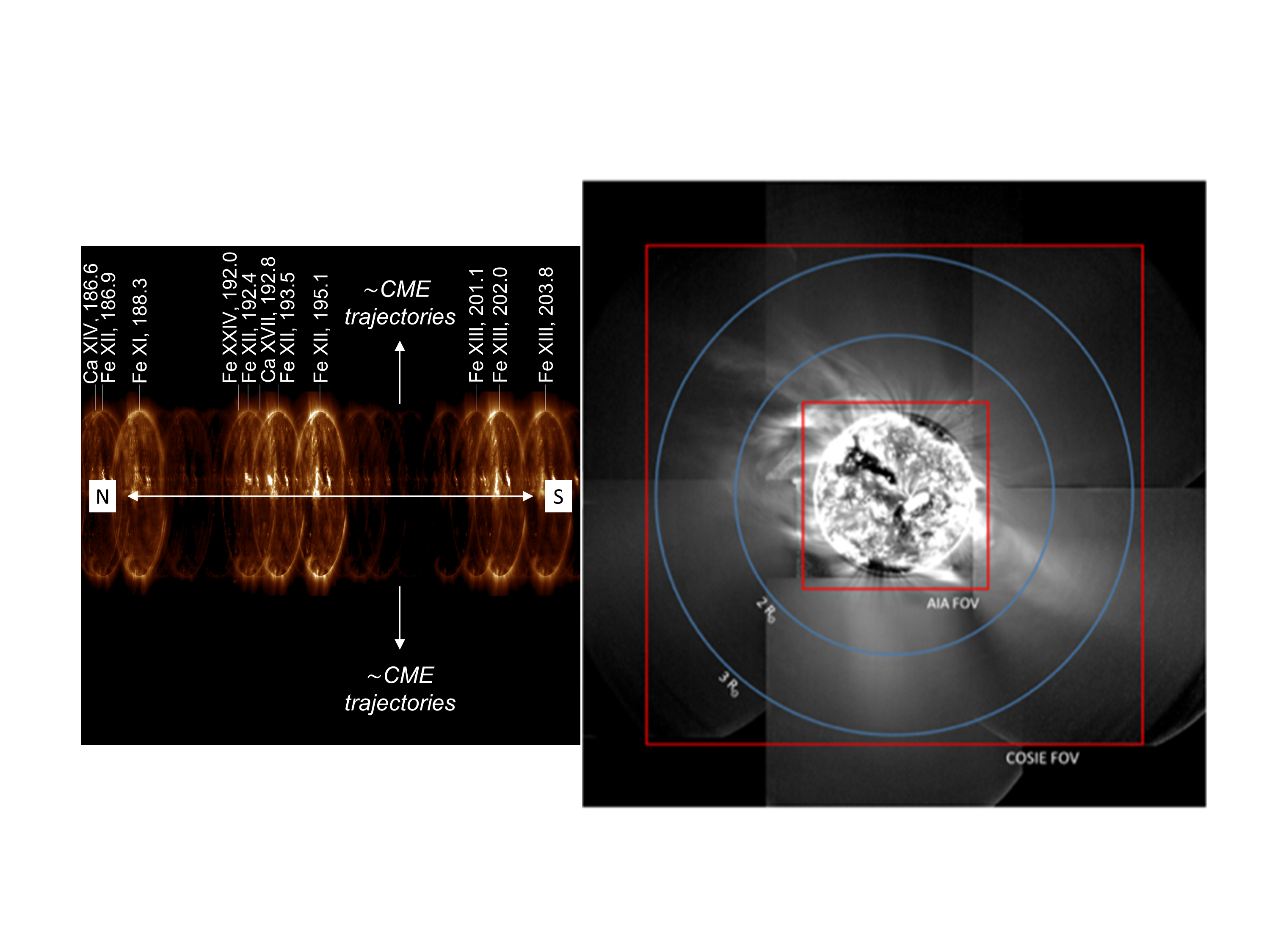}}
  \caption{An example of the expected COSIE data.  The left panel shows the COSIE-S overlapped spectral/spatial image. The N-S axis is in the horizontal direction so that CME trajectories will not be in the spectral direction.   The location of a few strong spectral lines are shown on the overlapped image.  They are printed at the spectral location associated with the solar central meridian.}  In the right panel, the COSIE-C FOV is shown on a SUVI wide-field mosaic. Additional examples of COSIE-S and -C data are given in Figures 4 and 14.
\label{fig:COSIE-C-S}
\end{figure}

Table~\ref{tab:linelist} gives the strongest lines in the COSIE-S wavelength range for each species, the temperature of maximum emissivity, and the expected signal in the COSIE-S spectroheliogram in photons s$^{-1}$ line$^{-1}$ 3.1\arc\ pixel$^{-1}$ for three representative cases: the quiet Sun, an active region and a flare.  To obtain the line intensities we used three corresponding emission measure files as available in the CHIANTI v.8 \citep{delzanna_etal:2015} database, noting that such estimates are approximate.   The strongest and brightest lines are Fe XII 19.5119\,nm and Fe XIII 20.2044\,nm, but there are lines from all Fe species from VIII-XIII in the COSIE-S wavelength range.  These provide the basis for the expected temperature sensitivity of COSIE (primarily 5.65 $\leq $ Log T $\leq$ 6.25).  Outside of this temperature range on the low end is the O V 19.2906\,nm line formed at Log T = 5.35.    Additionally, there are several weaker Ca lines that can provide temperature information from 6.55 $\leq $ Log T $\leq$ 6.75 for active regions. The Ca XVII 19.28 and Fe XXIV 19.204\,nm resonance lines, the hottest in the band, become very bright during solar flares and provide information on hotter plasma.  Note that the CHIANTI flare emission measure was obtained from a relatively large solar flare, a GOES class M2, during peak emission. For such large flares,  short exposure times would be required to avoid saturation.  This wavelength range also includes several density sensitive lines from Fe XII and Fe XIII; we include two of these in Table~\ref{tab:linelist}.   See \cite{young2007} and the review of \cite{delzanna_mason:2018} for additional spectral lines and diagnostics available in this wavelength range.

\begin{deluxetable}{c c | r r r}
\tablecaption{Strong Lines in the COSIE-S Wavelength Range \label{tab:linelist}}
\tablehead{
\colhead{Ion and}  & \colhead{Log Maximum} &\multicolumn{3}{c}{Expected Signal} \\
\colhead{Wavelength (nm)} & \colhead{Temperature} & \colhead{QS} & \colhead{AR}& \colhead{Flare}  }
\startdata
O V 19.2906 & 5.35  &  0.7  &  3.6  &  2116.5  \\
O VI 18.4117 & 5.45  &  0.6  &  7.4  &  503.9  \\
Fe VIII 18.5213 & 5.65  &  10.9  &  101.5  &  4460.3  \\
Fe IX 18.8497 & 5.85  &  10.1  &  104.3  &  1764.1  \\
Fe X 18.4536 & 6.05  &  18.8  &  234.1  &  2793.9  \\
Fe XI 18.8216 & 6.15  &  30.4  &  577.3  &  6568.7  \\
Fe XII 19.5119 & 6.20  &  30.8  &  1135.1  &  14355.0  \\
Fe XII 18.688 & 6.20$^a$  &  8.7  &  304.3  &  3804.5  \\
Fe XIII 20.2044 & 6.25  &  5.8  &  547.7  &  8618.1  \\
Fe XIII 20.3826 & 6.25$^a$  &  4.0  &  355.3  &  5516.6  \\
S XI 19.127 & 6.30  &  0.5  &  49.8  &  957.7  \\
Ar XIV 19.4396 & 6.55  &  0.0  &  8.1  &  1131.7  \\
Ca XIV 19.3874 & 6.55  &  0.0  &  20.6  &  3504.1  \\
Ca XV 20.0972 & 6.65  &  0.0  &  13.6  &  4838.3  \\
Ca XVII 19.2858 & 6.75  &  0.0  &  26.2  &  62204.1  \\
Fe XXIV 19.204 & 7.25  &  0.0  &  0.0  &  806844.8  \\

\enddata
\tablenotetext{a}{Density-sensitive line}
\tablecomments{The final three columns give the signal in the COSIE-S channel for three standard Chianti differential emission measures assuming a density of $10^9$ cm$^{-3}$ in ph s$^{-1}$ line$^{-1}$ 3.1\arc pixel$^{-1}$.}
\end{deluxetable}

\section{UNFOLDING METHOD}

To invert the spectrometer and coronagraph data, we follow the spectral decomposition method described in
the companion paper by Cheung et al. (2019). Their paper describes a general framework for performing
spectral decomposition and inversion on single slit (e.g. Hinode/EIS), multi-slit (e.g. the proposed MUSE mission), and instruments like COSIE using a sparse inversion technique similar to \cite{cheung2015}.

We first cast the problem as a set of linear equations, namely
\begin{equation}
   \mathbold{y_{obs}=Mx}
   \label{eq:matrix}
\end{equation}
where $\mathbold{y_{obs}}$ is an array that contains the COSIE S+C observational data, $\mathbold{x}$ is an array of emission measures and $\mathbold{M}$ is a matrix that describes how the emission measure maps into the detector for both the COSIE S+C channels.  We describe $\mathbold{M}$ for COSIE in Appendix~\ref{definem}.

When \cite{cheung2015} applied a sparse inversion method to AIA data, the emission measure distribution was only a function of temperature and the data matrix was the intensity in each AIA channel in a single pixel.  The returned emission measure distribution represented the emission measure along a single line of sight.  For COSIE-S, however, the emission measure along a single line of sight in COSIE-C maps onto multiple detector pixels in the spectral direction.  We treat each CCD row (along the dispersion direction) as an independent inversion problem. Hence, the emission measure distribution, $\mathbold{x}$, must at least be a distribution of both temperature and lines of sight that contribute to that row of COSIE-S data in the spectral direction. We solve for the emission measure distribution as a function of temperature (and perhaps other parameters) for all lines of sight that contribute to a single row of COSIE S+C data simultaneously. The COSIE-S channel has density sensitive spectral lines in the wavelength range, particularly from Fe XII and Fe XIII, and the COSIE resolution will be sensitive to strong ($> 50$ km s${-1}$) bulk flows.  Hence, the emission measure distribution as a function of density and velocity can also be considered. 

After establishing $\mathbold{M}$, we use the LASSOLARS routine to find the best solution for the emission measure distribution for a given row of COSIE S+C data. It is a Least Angle Regression \citep[LARS,][]{efron2004} implementation of the LASSO selection method \citep{tibshirani1996}. The routine is available in the Python \textit{scikit-learn} package \citep{scikit-learn}, we call it through an IDL-to-Python bridge. The underlying algorithm performs L1 regularization, i.e. it looks for a solution $\mathbold{x}^{\#}$ such that
\begin{equation}
    \mathbold{x}^{\#} = \mathrm{argmin} \Big[ {||\mathbold{y_{obs}-Mx}||^2_2} + \alpha||\mathbold{x}||_{1} \Big]
\end{equation}
where ${||\mathbold{y_{obs}-Mx}||^2_2}$ denotes the squared L2 norm of $(\mathbold{y_{obs}-Mx})$, this is a standard least squares expression. $||\mathbold{x}||_{1}$ is the L1 norm of $\mathbold{x}$ and the penalty that is applied to the solution. $\alpha$ is a hyperparameter that controls the degree of the penalty. In the case of $\alpha = 0$, the solution will be a standard least-squares fit. For $\alpha > 0$, the L1 penalty term in Equation 2 will get bigger the more different emission measures are required for the solution. In other words, it encourages \textit{sparse} solutions. As negative emission measure solutions are not physical, we only allow positive solutions.

To validate the approach and determine the optimal resolutions and ranges for the inversion, we apply this method to simulated COSIE data sets.  Unfortunately, there is not a single data set available that can be used to fully explore the all  parameters.   First we apply this method to data derived from a magnetohydrodynamic (MHD) model of the Sun; this is discussed in Section~\ref{mhdmodel}.  For the model, the temperature and density distribution are known along every line of sight.  Using this simulation, we determine the optimal spatial and temperature resolution of the inversion and demonstrate the method of determining the densities from the inverted line ratios.   The MHD model has a limited temperature range of plasma.  To explore how to incorporate different temperature ranges, we also apply this method to data derived from a full Sun emission measure calculation from AIA; this is discussed in Section~\ref{aiamodel}.  For the AIA model, the temperature distribution is known for every line of sight and there is emission at a broader range of temperatures than in the MHD model, but there is no density information.  We use the optimized spatial and temperature resolution derived in Section~\ref{mhdmodel} and explore how using different temperature ranges in the inversion alters the results.

\section{APPLYING THE UNFOLDING METHOD TO AN MHD MODEL \label{mhdmodel}}

One issue in solving the set of linear equations in Equation~\ref{eq:matrix} comes from redundant rows in the matrix $\mathbold{M}$, meaning when emission measure with different temperatures, densities, or line-of-sight (LOS) positions produce the same signature in the COSIE instrument.  For instance, the COSIE-S has a nominal spatial resolution of 9.3\arc\ in the spectral direction, though the thermal width of the spectral lines can degrade the spatial resolution even further.  If we define, then, two matrix rows for lines of sight that are only 1\arc\ apart, the resulting emission measure would produce essentially the same signal in COSIE-S.  Similarly, defining a row of $\mathbold{M}$ for a temperature or density where COSIE has little sensitivity returns no additional information and slows down the process for solving Equation~\ref{eq:matrix}.  The first step, then, is to determine the appropriate ranges and resolutions of the lines of sight, temperatures, and densities.  

We determine these best parameters piecemeal, meaning we first consider the best spatial and temperature resolution, then include density for those values only.  There are other parameters that can impact how well the set of equations can be solved that are beyond the inherent limitations of the instrument, such as the constant density or pressure we use if we are not solving for density or the elemental abundances in the emitting plasma.  Below we make several different assumptions with regards to the pressure and density and defer discussion of the potential impact of the abundances to Section \ref{sec:disc}.  
 
In this section, we use a three-dimensional MHD model of the solar atmosphere, developed by Predictive Science, Inc and described briefly below.  The model solution has temperature, plasma density, magnetic field strength and velocity defined at every point in the volume around the Sun. None of the velocities are large enough to be resolved by COSIE-S, so for now we ignore the potential impact of the velocity on  our inversion. We revisit this choice in Section \ref{sec:disc}.  

\subsection{Description of the Model}

To ensure reasonably realistic values and spatial variation of temperature and density in the solar corona, we use a high-resolution calculation by the 3D Magnetohydrodynamic Algorithm outside a Sphere (MAS) code \citep[e.g.][]{mikic99,lionello09,downs13,caplan17}, which was used to predict the structure of the global corona during the 21 August 2017 total solar eclipse. Described in detail by \citet{mikic18_tmp}, this calculation employed high-fidelity magnetic field observations at the inner boundary, a new Wave-Turbulence-Driven (WTD) coronal heating model \citep{downs16}, and a method to energize large-scale magnetic flux-systems in the low corona. Forward modeled observables, including SDO/AIA EUV images, temperature maps from DEM inversions, and broadband visible light images, compared favorably with observations, making this an ideal model to use here. The minimum and maximum temperatures of the simulation range from 0.01 to 3~MK. However, the transition region is artificially broadened to resolve the self-regulation of mass and energy in the low corona \citep[$T_c$ = 0.35~MK, see][]{lionello09,mikic13}, and we only use temperatures above 0.4~MK in this analysis.

Based on the COSIE-C resolution and FOV, we first define a grid of lines of sight in a square that is $\pm$ 3 solar radii from sun center at a resolution of 3.1\arc.  We then extract the emission measure along each line of sight as a function of both density and temperature.  We fold this emission measure distribution through the COSIE response matrix and generate the simulated COSIE S+C detector images row by row. Recall that along a CCD row in the spectral dispersion direction, COSIE-C has a spatial resolution of 3.1\arc per pixel and COSIE-S has a spatial resolution of 9.3\arc per pixel. 

The nominal COSIE exposure time will be 1 second, so we multiply by this exposure time and then get photons pixel$^{-1}$.  We add Poisson noise to each pixel and then average 20 noisy frames (approximately one minute of observations) together to mimic the expected data.  Figure~\ref{fig:cosie_sc_truth} shows these images for the central portion of the detector.  

\begin{figure}[ht]
    \centering
    \resizebox{.8\textwidth}{!}{\includegraphics{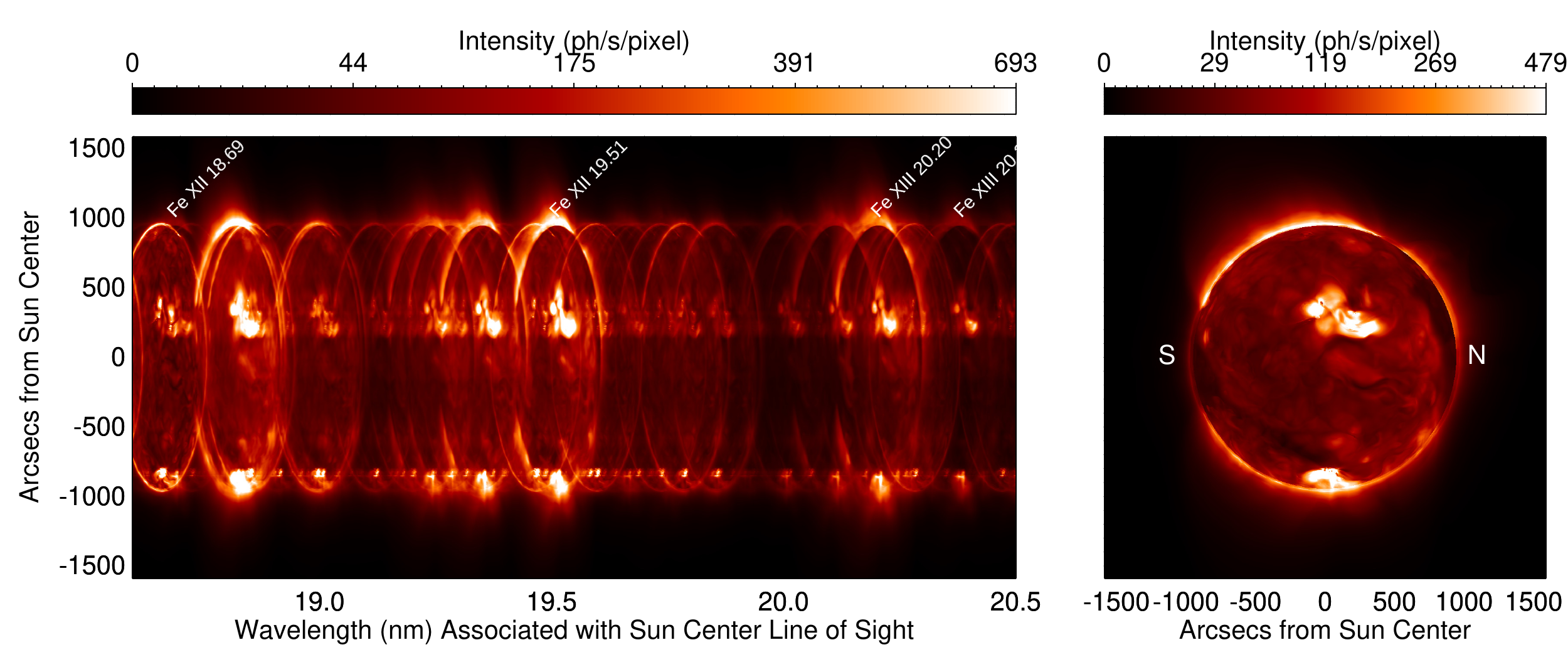}}
    \caption{The COSIE spectroheliogram (left) and coronagraph images associated with the output of the MHD model.  Only the central portion of the detector is shown.  The assumed exposure time is 1 s, noise has been added and 20 images have been averaged.  The Sun is oriented with North to the right in the images.}
    \label{fig:cosie_sc_truth}
\end{figure}

\subsection{Determining the optimal spatial and temperature resolution}

First, we consider only an inversion to determine the temperature distribution of the emission measure.  We calculate $M$ for constant pressures of $10^{15}$ and $10^{16}$ K cm$^{-3}$ and constant densities of $10^8$ and $10^9$ cm$^{-3}$.  As input, we use the COSIE data simulated from Predictive Science Model (see Figure~\ref{fig:cosie_sc_truth}).  Note that this data includes density dependence in spectral lines even though we ignore the density dependence in this initial inversion.  

We perform inversions with a temperature resolution of $\Delta$ Log T of 0.1 and 0.2.   The temperature range is limited to 5.8 $<$ Log T $<$ 6.4 for all of the inversions.  We perform the inversions with three different spatial resolutions, $\Delta s$ of 9.3\arc, 18.6\arc, and 27.9\arc.   For all of the inversions we limit the contributing FOV to $\pm1.2$ solar radii.  This implies that the code only has the option to put emission measure in spatial bins that are $\Delta s$ wide that extend to $\pm1.2 R_\odot$ in the spectral (N-S) direction.  The inversion code is run on each row of data independently so the final emission measure map will have a resolution of $\Delta s \times $3.1\arc.  Also, there is no limit on the spatial rows that can be calculated, meaning the inversion can be run on rows at distances in the E-W direction larger than $1.2 R_\odot$ if there is adequate signal at those distances.  In this test case, we compute the inversion for 1000 rows of the CCD, which is equivalent to $\pm 1.6 R_\odot$ in the solar E-W direction.  A description of the parameters used in this initial inversions and a summary of the results are given in Table~\ref{tab:tempmap}.

For each set of parameters, we predict the inverted COSIE-S and -C data.  We calculate the reduced $\chi^2$ by comparing the original full resolution COSIE-S data with  the inverted COSIE-S data.  This is also given in Table~\ref{tab:tempmap}. Note that the number of degrees of freedom (DoF) are different for each row of Table~\ref{tab:tempmap}.  The larger the DoF, the smaller the normalization factor when calculating the reduced $\chi^2$.   For instance, the ratio of the normalization factor between T15 (with 9.3\arcsec spatial resolution) and T16 (with 18.6\arcsec spatial resolution) is a factor of 3.8, meaning if the inverted data was, in all other ways, equivalent between the two, the reduced $\chi^2$ for T15 would be 3.8 times the reduced $\chi^2$ for T16.  The difference in the $\chi^2$ values between these two inversions is roughly a factor of 40, indicating that the inversion itself is worse in the T15 case.

\begin{deluxetable}{cccccc}
\tablecaption{Summary of Inversion Parameters to Determine Temperature Maps \label{tab:tempmap}}
\tablehead{
\colhead{Run} & \colhead{LOS} & \colhead{}  & \colhead{Log Constant}  & \colhead{Log Constant}  & \colhead{$\chi^2$}  \\
\colhead{Number} & \colhead{Resolution} & \colhead{$\Delta$ Log T} & \colhead{Pressure}  
 & \colhead{Density} & \colhead{COSIE-S}\\
\colhead{} & \colhead{(arcsec)} & \colhead{} & \colhead{(K cm$^{-3}$)} &  \colhead{cm$^{-3}$)} &  \colhead{} 
}
\startdata
T0   &   9.3  &  0.2  &  15  &  N/A  &  10.1  \\  
T1   &   18.6  &  0.2  &  15  &  N/A  &  5.8  \\  
T2   &   27.9  &  0.2  &  15  &  N/A  &  5.3  \\  
T3   &   9.3  &  0.1  &  15  &  N/A  &  38.9  \\  
T4   &   18.6  &  0.1  &  15  &  N/A  &  5.9  \\  
T5   &   27.9  &  0.1  &  15  &  N/A  &  4.7  \\  
T6   &   9.3  &  0.2  &  16  &  N/A  &  34.4  \\  
T7   &   18.6  &  0.2  &  16  &  N/A  &  24.0  \\  
T8   &   27.9  &  0.2  &  16  &  N/A  &  21.7  \\  
T9   &   9.3  &  0.1  &  16  &  N/A  &  87.0  \\  
T10   &   18.6  &  0.1  &  16  &  N/A  &  23.2  \\  
T11   &   27.9  &  0.1  &  16  &  N/A  &  18.6  \\  
T12   &   9.3  &  0.2  &  N/A  &  8  &  6.4  \\  
T13   &   18.6  &  0.2  &  N/A  &  8  &  1.1  \\  
T14   &   27.9  &  0.2  &  N/A  &  8  &  1.1  \\  
T15   &   9.3  &  0.1  &  N/A  &  8  &  37.8  \\  
T16   &   18.6  &  0.1  &  N/A  &  8  &  1.3  \\  
T17   &   27.9  &  0.1  &  N/A  &  8  &  0.8  \\  
T18   &   9.3  &  0.2  &  N/A  &  9  &  14.2  \\  
T19   &   18.6  &  0.2  &  N/A  &  9  &  8.7  \\  
T20   &   27.9  &  0.2  &  N/A  &  9  &  7.9  \\  
T21   &   9.3  &  0.1  &  N/A  &  9  &  46.1  \\  
T22   &   18.6  &  0.1  &  N/A  &  9  &  9.3  \\  
T23   &   27.9  &  0.1  &  N/A  &  9  &  7.3  \\  

\enddata
\tablecomments{All calculations are made for a spatial range (N-S) of $\pm$ 1.2 solar radii and a temperature range of $5.8 \leq $ Log T $\leq 6.4$.}
\end{deluxetable}

The lowest values of $\chi^2$ are associated with inversions completed with a constant density of $10^{8}$ cm$^{-3}$ (see runs T12 - T17).  This likely reflects that the densities in the model are close to $10^8$ cm$^{-3}$, particularly at the temperatures where there are the density sensitive lines in the COSIE wavelength range.   When comparing these 6 runs, the ones with a spatial resolution of 9.3\arc\  (T12 and T15) have a higher $\chi^2$. For this reason, we discard this resolution and focus on the 18.6\arc and 27.9\arc\ resolutions (T13, T14, T16, T17).   The $\chi^2$ for these four inversions are essentially identical and $\sim 1$.  In the interest of maintaining the highest spatial and temperature resolution possible, we  choose to continue with a temperature resolution of $\Delta$ Log T = 0.1 and a spatial resolution of 18.6\arc. 

Note these optimal resolutions are  highly dependent on the COSIE instrument design.  Nominally COSIE-S has 9.3\arc\ spatial resolution in the spectral direction, but because spectral lines are thermally broadened, this degrades the spatial resolution in the spectral direction further.   The COSIE wavelength range contains spectral lines from Fe VIII - XIII, which provides excellent temperature discrimination in this temperature range.

\subsection{Including density in the inversion}

Next, we include density in the inversion, but we do not allow density to be a free parameter for all lines of sight or temperature bins.  Instead we allow density to be a free parameter only for lines of sight that have greater than a specific signal in the coronagraph.  We investigate how changing this cutoff impacts the inversion.  Along all other lines of sight, where the coronagraph intensity is less than this cut off, we use the constant density assumption with Log n = $10^{8}$ cm$^{-3}$.   We also apply a similar criterion for temperature.  We only allow for density to be a free parameter for the 6.1 $\leq$ Log T $\leq$ 6.3 temperature bins.  For all other temperature bins, we use the constant density assumption.   The results of the parameter study are summarized in Table~\ref{tab:densmap}.   

\begin{deluxetable}{ccccc}
\tablecaption{Summary of Inversion Parameters to Determine Density Maps \label{tab:densmap}}
\tablehead{
\colhead{Run}    & \colhead{Signal in}  & \colhead{$\chi^2$} & \colhead{Number of} \\
\colhead{Number} & \colhead{COSIE-C}   & \colhead{COSIE-S}  & \colhead{Rows in }\\
\colhead{}      & \colhead{(ph s$^{-1}$ pixel$^{-1}$)}  & \colhead{}  & 
\colhead{Combined Data}
}
\startdata
D0   &   100  &  12.4  &  749  \\  
D1   &   200  &  15.4  &  177  \\  
D2   &   300  &  4.1  &  56  \\  
D3   &   400  &  2.5  &  19  \\  
\multicolumn{3}{c}{Combined}  &  0.8  &   NA  \\  

\enddata
\tablecomments{All calculations are made for a spatial range of $\pm$ 1.2 solar radii, a LOS resolution of 18.6\arc, a temperature range of $5.8 \leq $ Log T $\leq 6.4$, and a temperature resolution of $\Delta$ Log T = 0.1.  The density is only allowed to vary in LOS positions where the signal in COSIE-C is greater than the value in the second column  and in the temperature bins 6.1 $\leq$ Log T $\leq$ 6.3. For all other LOS positions and temperature bins, we assume a constant density at $10^8$ cm$^{-3}$.  The final column gives the number of rows from each of the individual runs that go into the combined solution.    The final row is the  $\chi^2$ associated with a combined data set.}
\end{deluxetable}

Ideally, we would use the solution associated with the fewest restrictions, meaning the lowest signal in the coronagraph or Run D0 in Table~\ref{tab:densmap}.  However, the $\chi^2$ associated with the full Sun inversion is high (12.4).  This is because for some rows of the inversion, we do not find an acceptable solution with this signal cutoff. These failed rows dominate the $\chi^2$. Fortunately, for those rows, we can simply use the solution from a different run that finds an acceptable solution.  Hence we combine the inversions from all the runs into a final solution by evaluating each row of the inversion individually.  

To explain, we show the $\chi^2$ associated with three individual rows of the inversion in Figure~\ref{fig:chi2_row_example}.  The Run Number plotted on the x-axis corresponds to the Run Number given in Table~\ref{tab:densmap}. The horizontal dashed line shows the minimum $\chi^2$ for all runs for that specific row times 1.5.  We assume any solution that falls below this $\chi^2$ value is acceptable.  We select the solution with the most free parameters (lowest run number) that meet this criterion for that row and store the emission measure distribution as a function of line of sight, temperature and density for that row in a master array. For the example row shown in the top plot, there is a high $\chi^2$ for Run D0, but a low $\chi^2$ for Run D1.  For this row in the combined solution, we use the solution from D1.  For the middle example row, all the runs have the same $\chi^2$ so we use the solution from D0.  For the bottom example row, the D0 solution has the lowest $\chi^2$, so we use it in the combined solution.   

\begin{figure}[ht]
    \centering
    \resizebox{.5\textwidth}{!}{\includegraphics{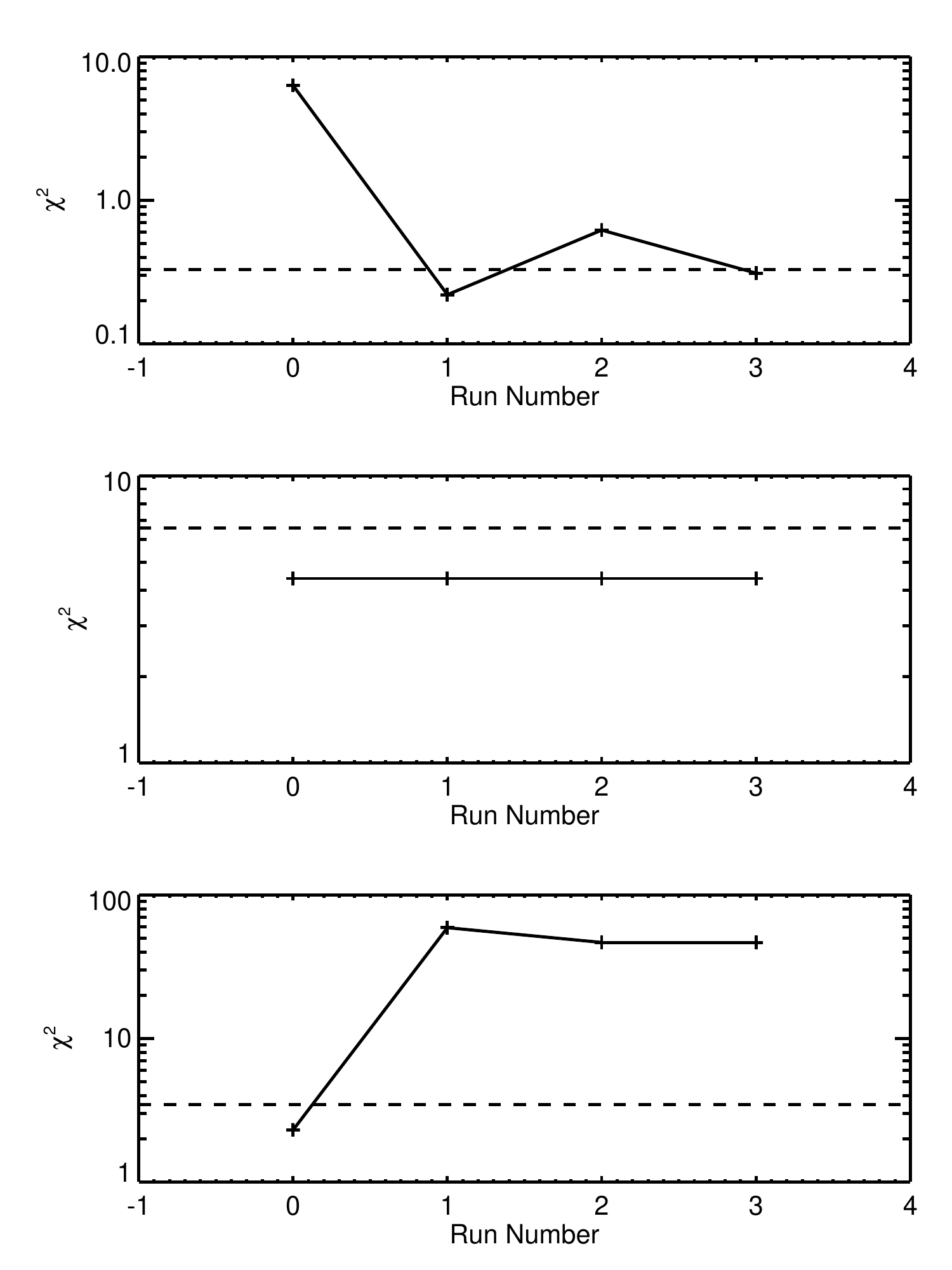}}
    \caption{$\chi^2$ for the four runs given in Table~\ref{tab:densmap} for three example rows of the inversion.  The horizontal dashed line shows 1.5 times the minimum $\chi^2$.  We assume any value of $\chi^2$ less than this value is acceptable and choose acceptable solution with the most free parameters (the lowest run value) for that row.}
    \label{fig:chi2_row_example}
\end{figure}

In this way, row by row, we build up the best estimate of the emission measure distribution for the entire field of view. The final column of Table~\ref{tab:densmap} gives the number of rows of data that come from each run number.  The majority or rows come from Run D0 and D1.  Note the combined solution has a normalized $\chi^2$ of $0.8$.  The resulting best spectrometer and coronagraph image associated with the combined data set is shown in the top panels of Figure~\ref{fig:cosie_sc_inv}.  The bottom panels show a difference image between these data and the input data from the simulations shown in Figure~\ref{fig:cosie_sc_truth}. 

\begin{figure}[ht]
    \centering
    \resizebox{.8\textwidth}{!}{\includegraphics{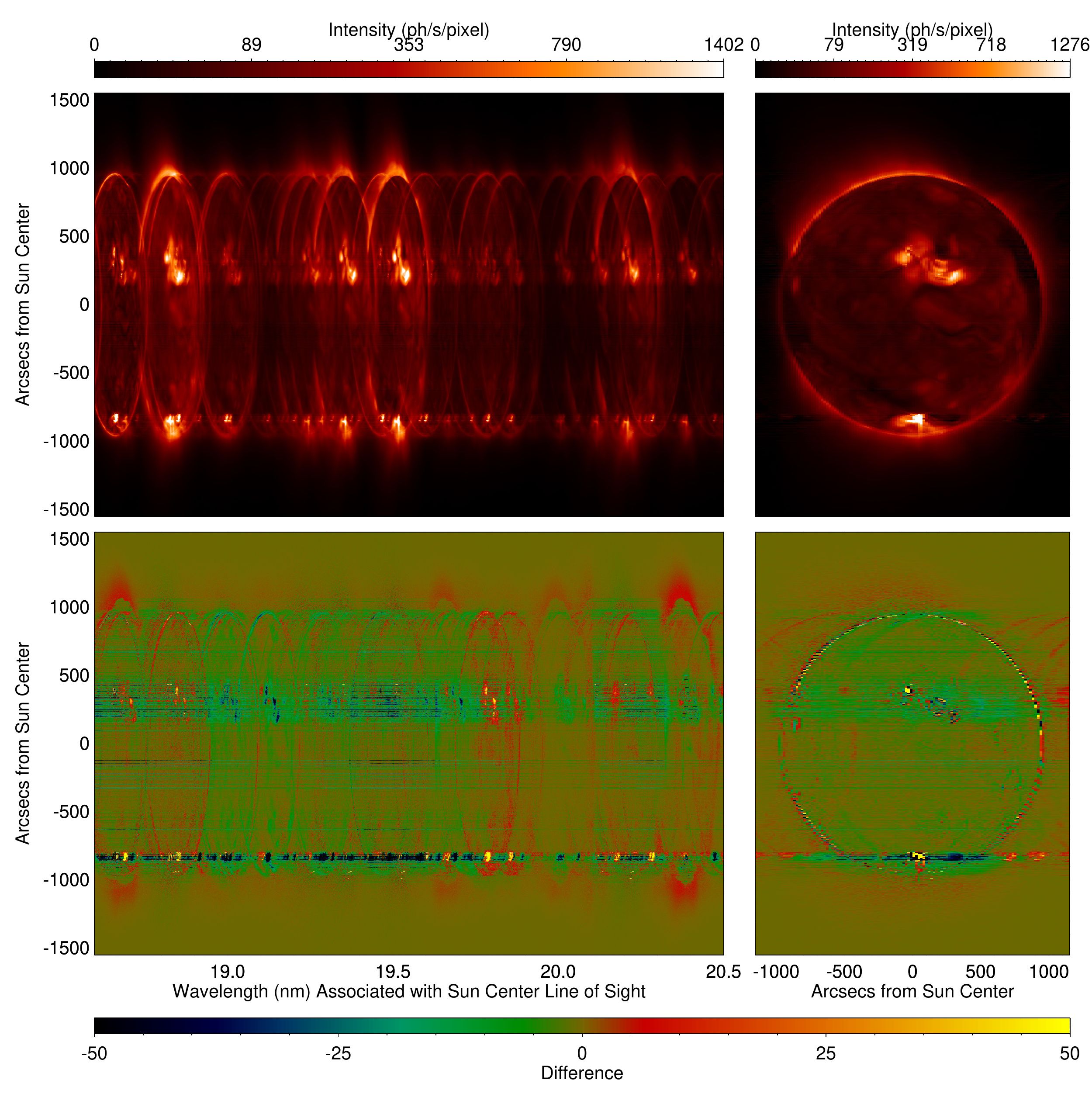}}
    \caption{The spectrometer and coronagraph data (top panels) associated with the best inversion scaled to the 0.5 power.  The difference between these data and the input data in Figure~\ref{fig:cosie_sc_truth} are shown in the bottom panels.}
    \label{fig:cosie_sc_inv}
\end{figure}

\subsection{Comparing true and inverted data}

In the above subsections, we describe selecting the best parameters for the inversion.   We made the selection of the best parameters based only on comparing the input and inverted spectrometer data.  This is identical to how we would treat real observations as well, we would only have access to the observations to perform and evaluate the inversions.  In this subsection, we compare the resulting temperature-emission measure maps, spectrally pure intensities, density maps, and spectra with those of the truth data.

Figure~\ref{fig:tempmap_with_dens} shows the true emission measure map as a function of temperature from the Predictive Science model in the first and third columns and the inverted emission measure map in the second and fourth columns.   Agreement is particularly good at temperatures that are well covered by strong lines given in Table~\ref{tab:linelist}.

\begin{figure}[ht]
    \centering
    \resizebox{.8\textwidth}{!}{\includegraphics{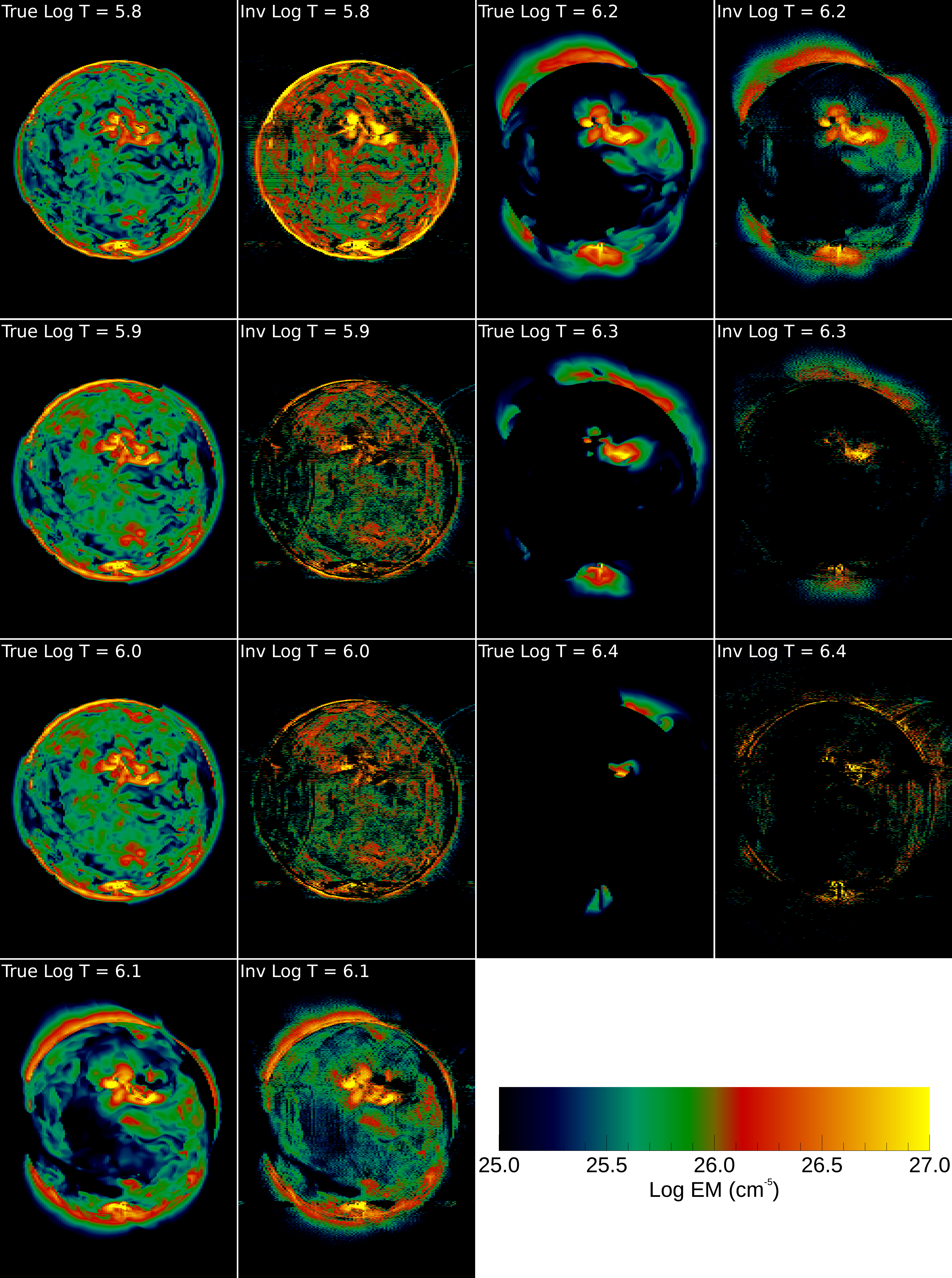}}
    \caption{The true emission measure distribution in different Log T bins 5.8-6.4 is shown in the first and third columns, the inverted EM distribution in the same temperature bins is shown in the second and fourth columns.}
    \label{fig:tempmap_with_dens}
\end{figure}

Next we calculate the spectrally pure line intensities in the strong Fe XII and XIII lines given in Table~\ref{tab:linelist}, including two density sensitive lines.  These comparisons are shown in Figure~\ref{fig:comp_moments_fe12}.  The images show the intensity maps for the true and inverted spectral lines.  The line plots show histograms of percentage error for all the pixels where more than 30 photons s$^{-1}$ pixel$^{-1}$ are expected. The average and standard deviation of the percentage errors are given in the histogram plots.  We calculate the standard deviation of the percentage error and find that for the Fe XII 19.5\,nm line, the standard deviation is 15\%, for the Fe XII 18.6\,nm line, the standard deviation is 35\%.  For the Fe XIII 20.2\,nm line, the standard deviation is 21\%, and for the Fe XIII 20.3\,nm line, the standard deviation is 30\%.  

\begin{figure}[!ht]
    \centering
    \resizebox{.8\textwidth}{!}{\includegraphics{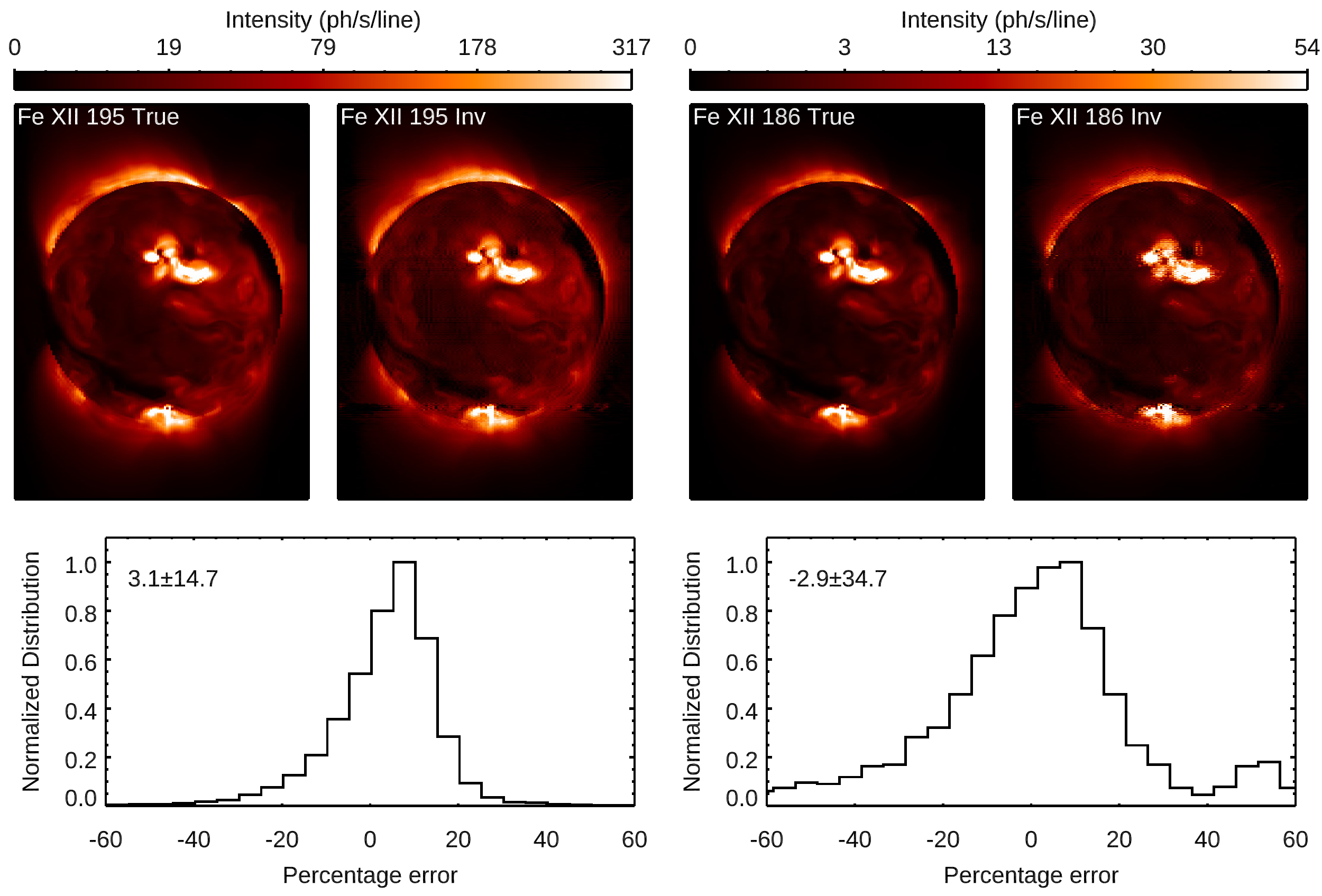}} 
    \resizebox{.8\textwidth}{!}{\includegraphics{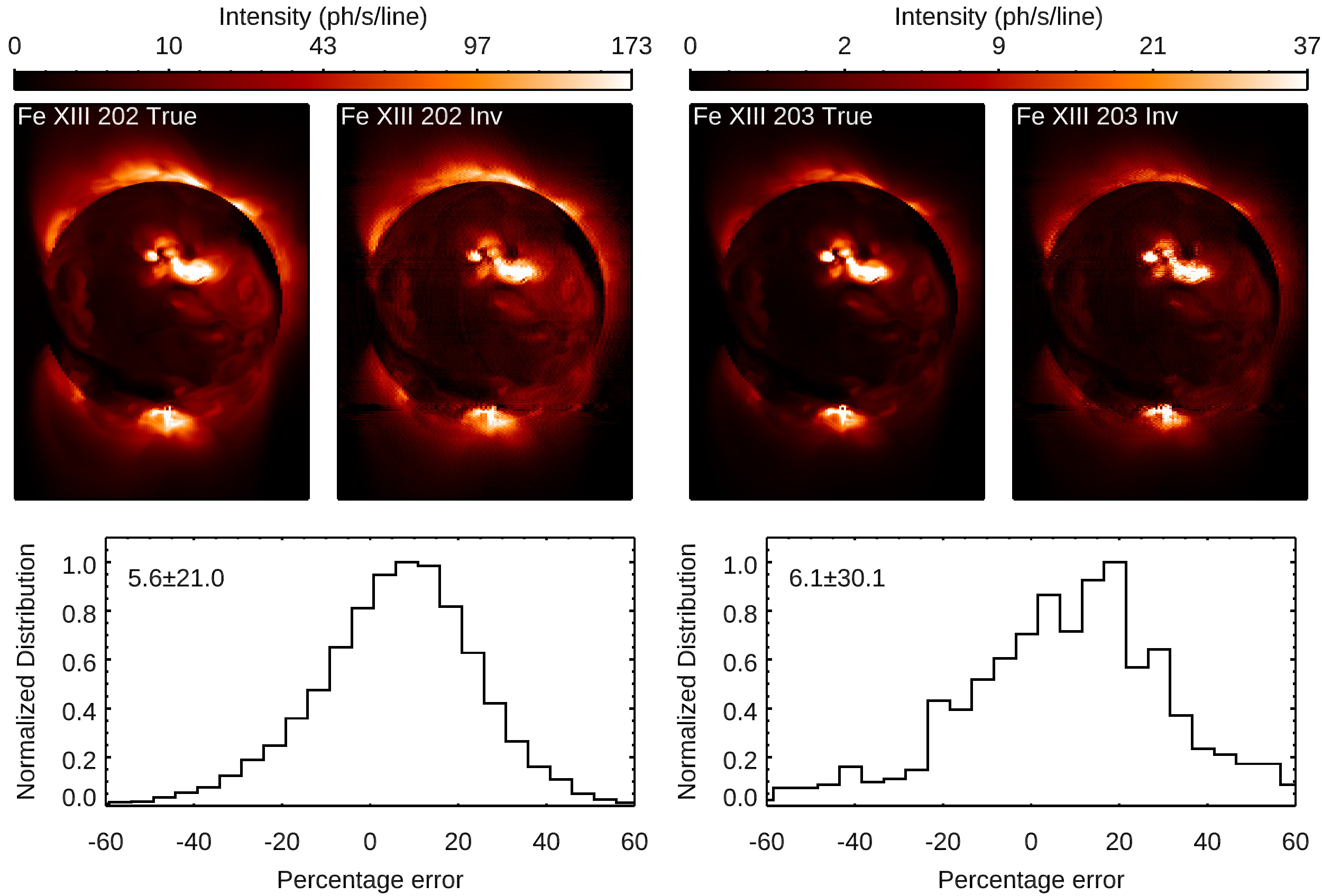}}
    \caption{A comparison of the true and predicted Fe XII (top) and Fe XIII (bottom) lines.  Intensity maps are scaled to the 0.5 power.  The histograms reflect the percentage errors for all pixels with greater than 30 photons s\none\ line\none.  The average and standard deviation of the percentage error is given in each line plot.
    \label{fig:comp_moments_fe12}}
\end{figure}

We use the Fe XII and XIII line ratios to calculate the densities.  We use the intensities from the true ratios and the inverted ratios.  Full sun maps of the true and inverted ratios and the densities derived from them are shown in Figure~\ref{fig:dens_ratio}.  In these maps, we have masked the ratios and the densities where the intensity in the density sensitive line (Fe XII 18.6\,nm or Fe XIII 20.3\,nm) is less than 10 photons s\none line\none.  [Figures with different mask levels are available in the on-line version of this article.] 

In Figures~\ref{fig:densmap12_1_ar} and \ref{fig:densmap13_1_ar}, we show the ratios and densities in the area around the on-disk active region.  In these images, we applied a mask of 1 photon s\none line\none in the density sensitive lines.  In some pixels at the edges of the active region, the ratio and density in the inverted data differ significantly from the ratio and density of the true data.  In these regions, the line intensity is small.  The plots on the right hand side of the figures show the error of the log of the density as a function of the intensity in the density sensitive line and the cumulative distribution of the error in the log density for different values of intensity. We show that the error depends on the intensity in the density sensitive line, but in general less than 0.2.

\begin{figure}[ht]
    \centering
    \resizebox{.8\textwidth}{!}{\includegraphics{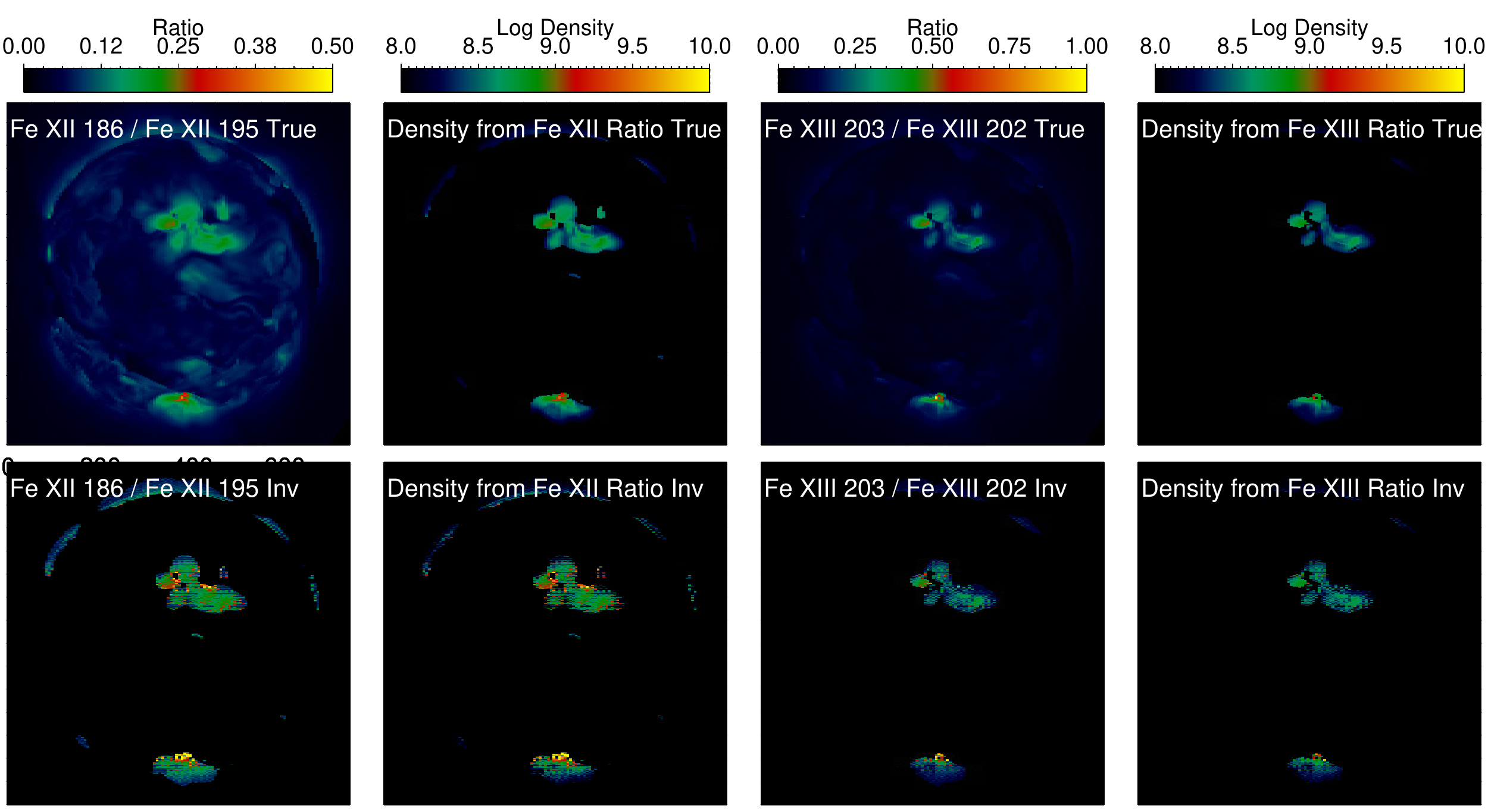}}
    \caption{The full sun ratio of the Fe XII and XIII lines from truth data and from inverted data, as well as the density determined from the ratio.  A mask has been applied to the inverted ratios, and all densities setting the values to 0 if the density sensitive line is less than 10 photons s\none line\none.  This image is available with other mask levels electronically.  The on disk active region is shown in Figures~\ref{fig:densmap12_1_ar} and \ref{fig:densmap13_1_ar}.}
    \label{fig:dens_ratio}
\end{figure}

\begin{figure}[ht]
    \centering
    \resizebox{.8\textwidth}{!}{\includegraphics{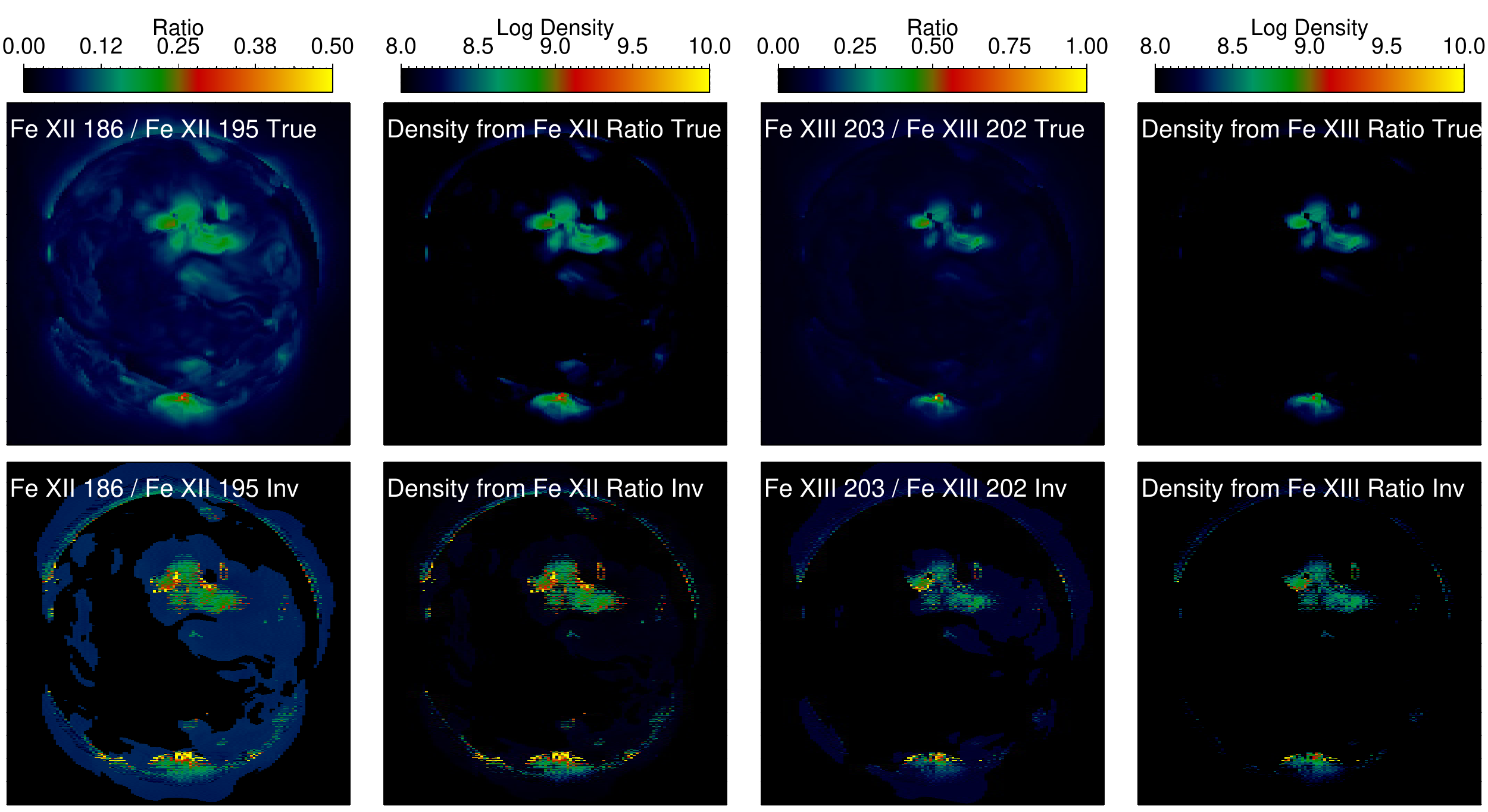}}
    \caption{Same as Figure~\ref{fig:dens_ratio} but with mask level of 1 photon s\none line\none.  In the paper, this will be included in electronic version only.}
    \label{fig:dens_ratio_1}
\end{figure}

\begin{figure}[ht]
    \centering
    \resizebox{.8\textwidth}{!}{\includegraphics{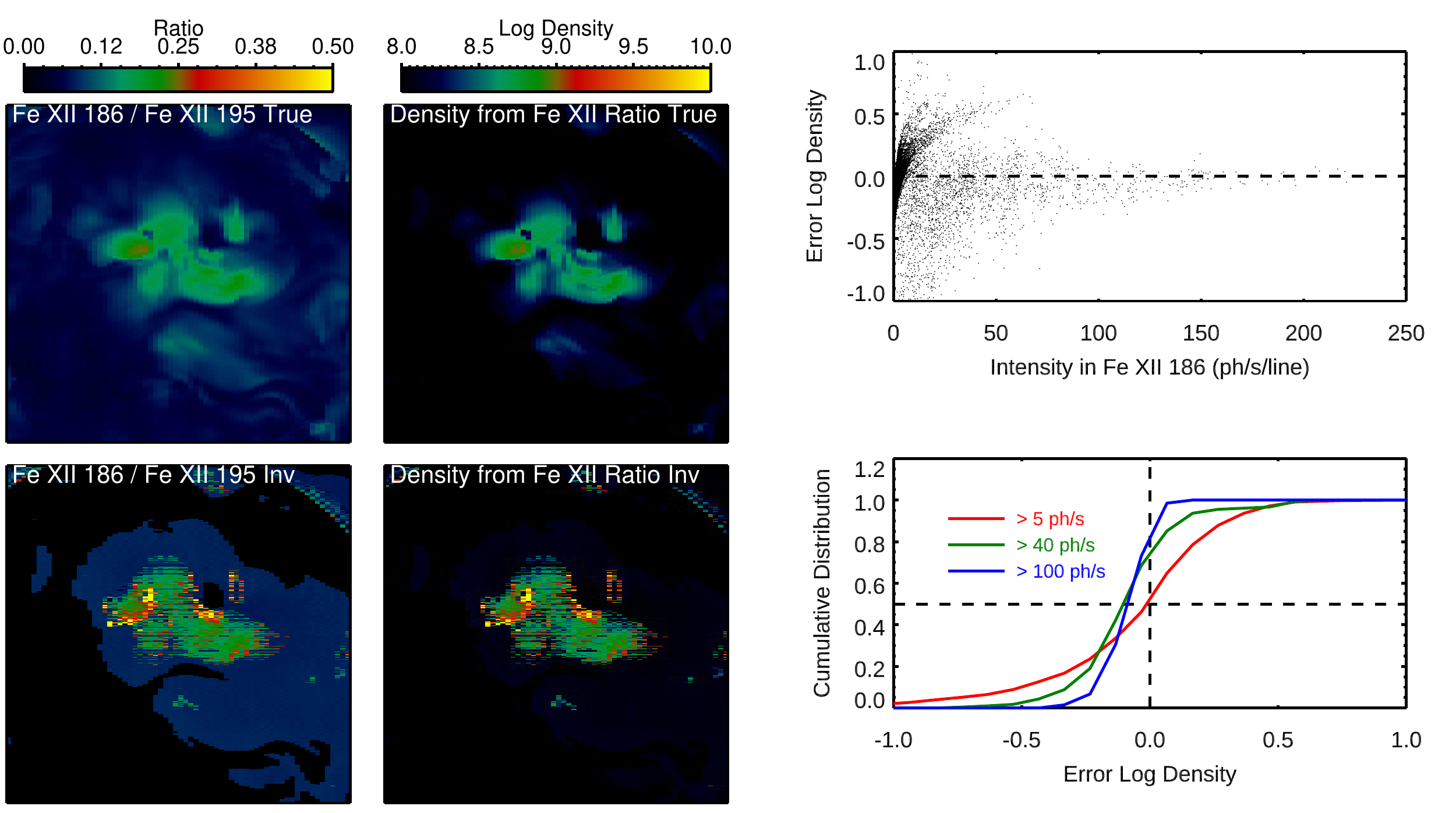}}
    \caption{Fe XII active region density map for the on disk active region with 1 photon s\none line\none mask.  The true ratio and density is shown in the top panels, the inverted are shown in the bottom panels.  The two plots on the right compare the true and inverted densities.  The top plot shows the error in the log of the density as a function of the intensity in the Fe XII 18.6\,nm line.  The bottom plot shows the cumulative distribution of the errors in the log of the density.  The red curve is all pixels with intensity larger than 5 photon s\none line\none, the green is 40 photons s\none line\none, and the blue 100 photons s\none line\none in the density sensitive line. }
    \label{fig:densmap12_1_ar}
\end{figure}

\begin{figure}[ht]
    \centering
    \resizebox{.8\textwidth}{!}{\includegraphics{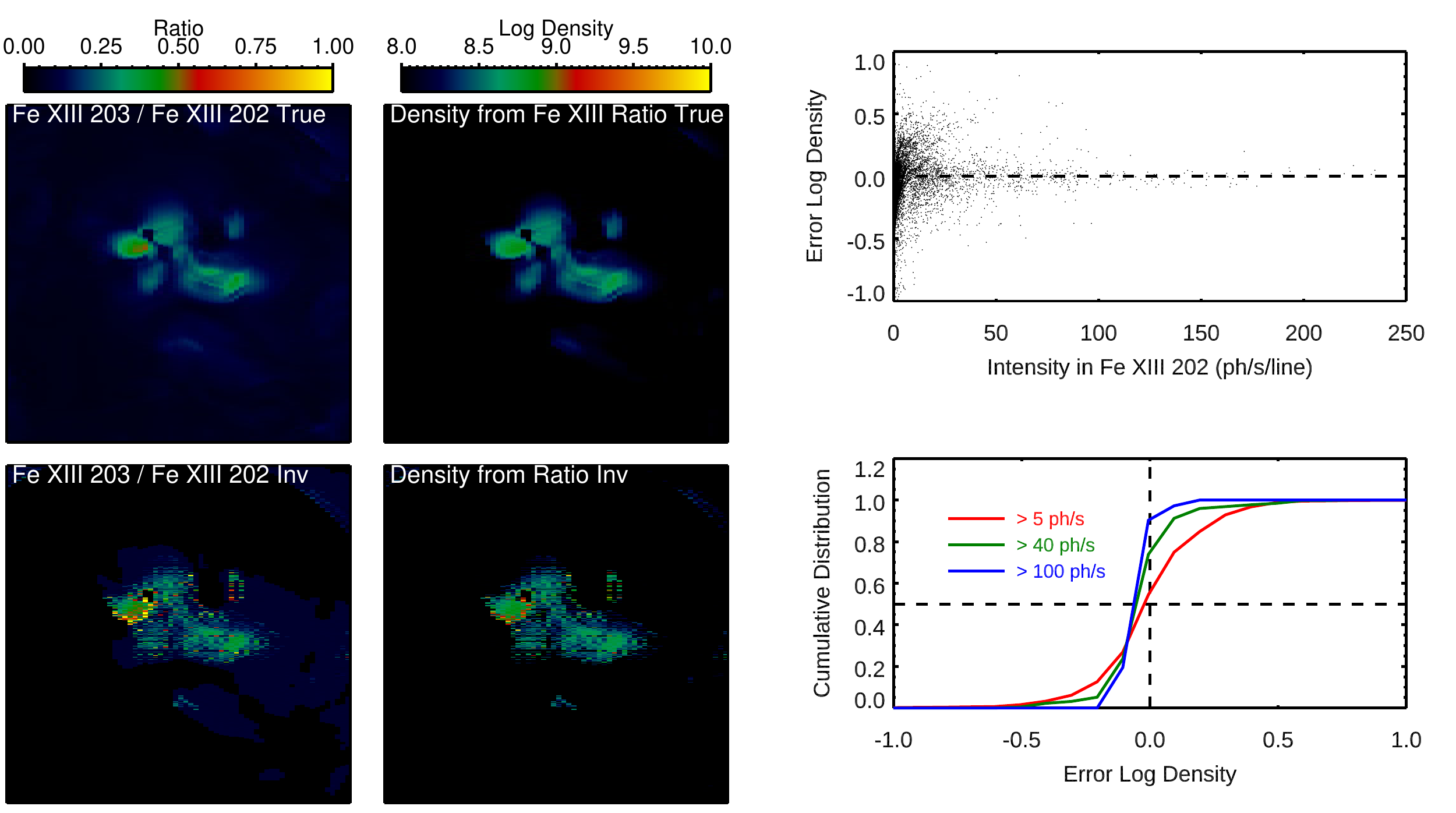}}
    \caption{Fe XIII active region density maps.  See caption for Figure~\ref{fig:densmap12_1_ar}.
    \label{fig:densmap13_1_ar}}
\end{figure}

Finally, we extract the full spectra along a single line of sight and compare the true to the inverted spectra.  A comparison of an active region, quiet sun, and limb spectra is shown in Figure~\ref{fig:example_spectra}.   

\begin{figure}[ht]
    \centering
    \resizebox{.8\textwidth}{!}{\includegraphics{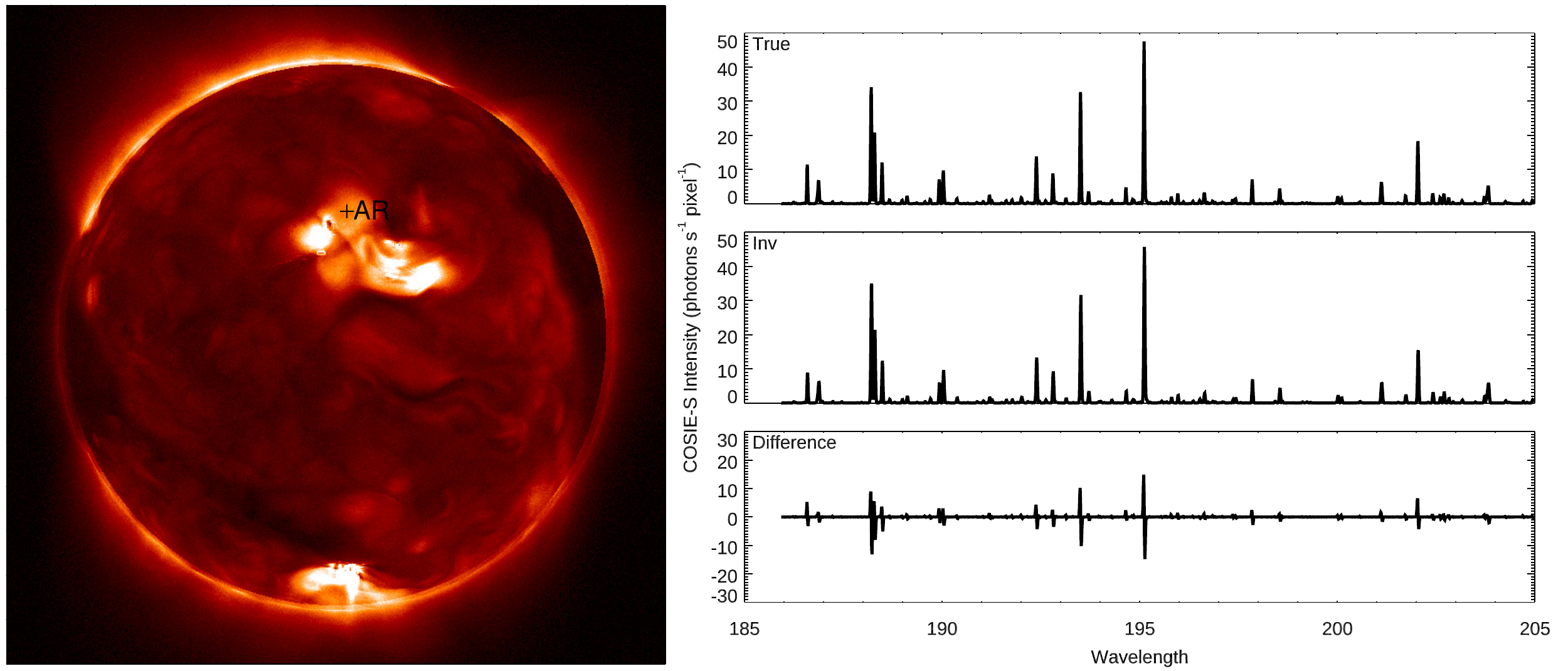}}
        \resizebox{.8\textwidth}{!}{\includegraphics{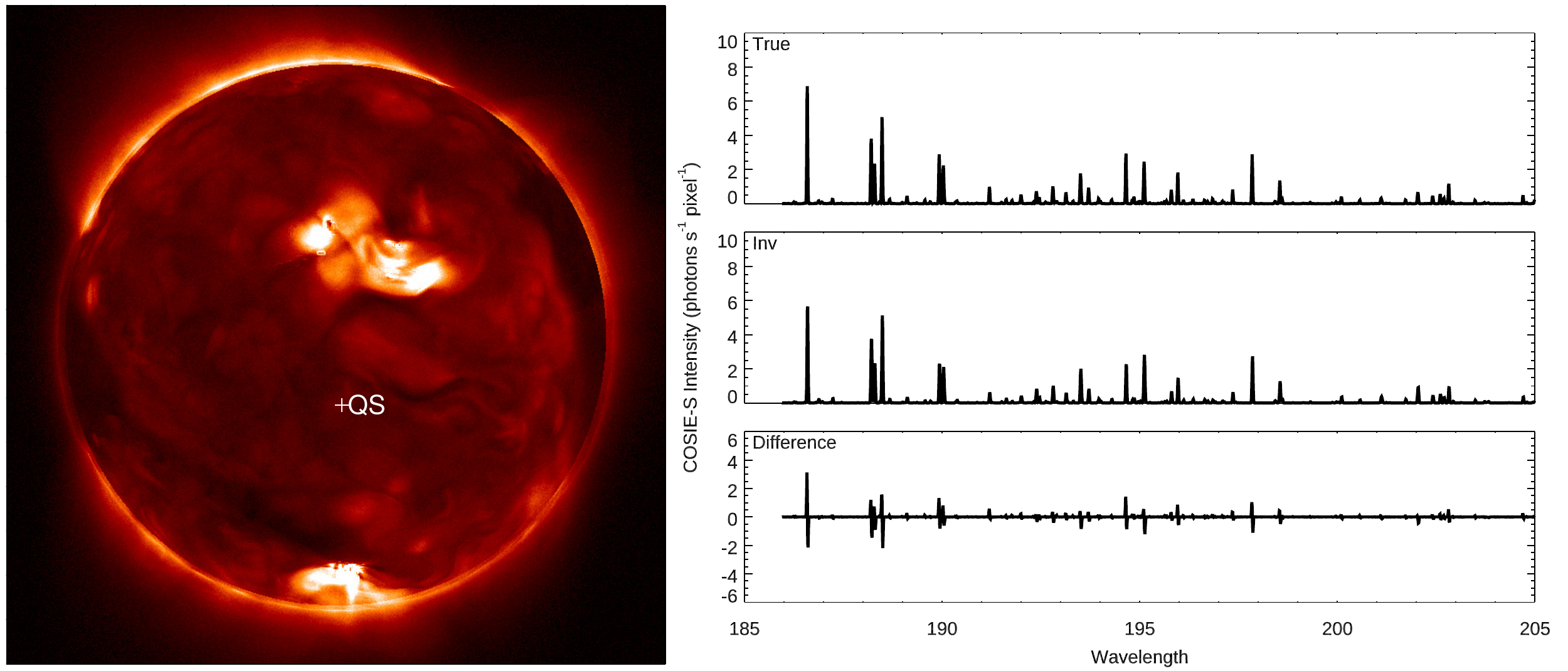}}
            \resizebox{.8\textwidth}{!}{\includegraphics{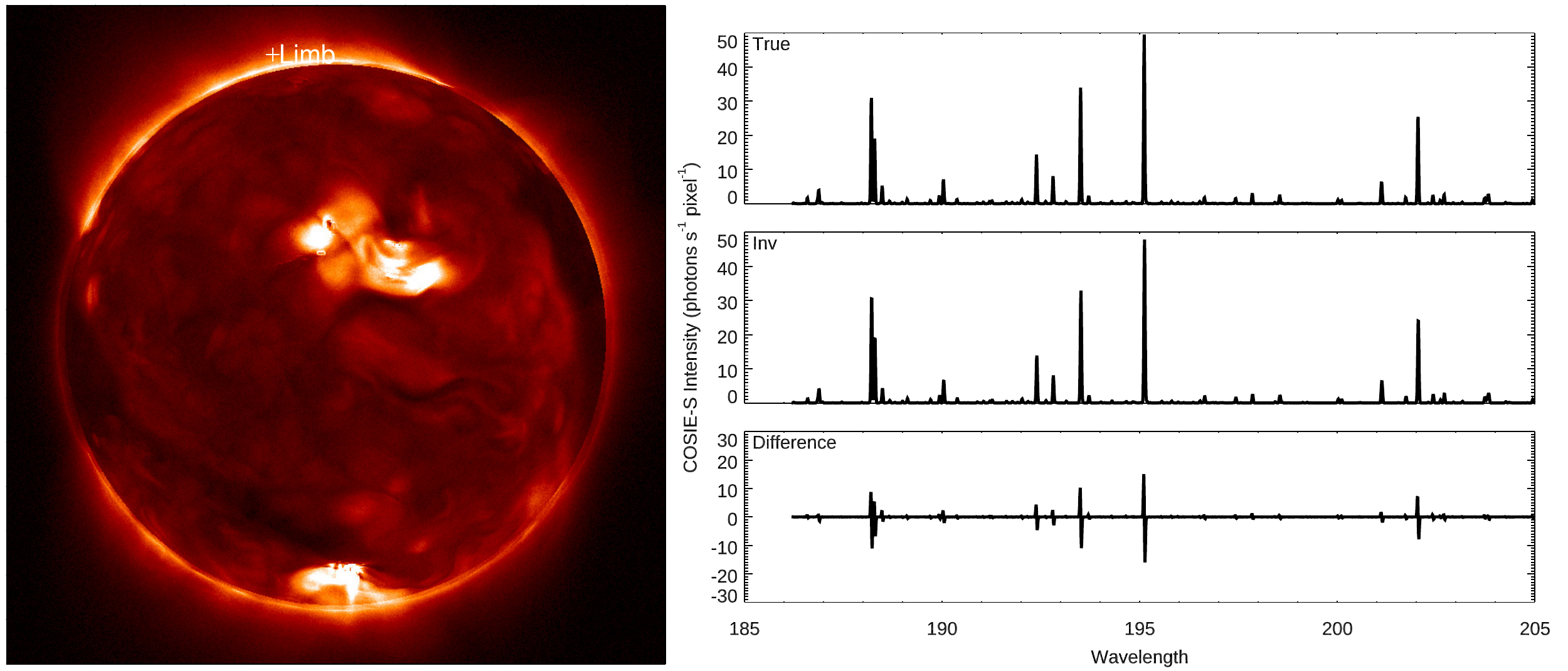}}
    \caption{An example of active region, quiet sun, and limb spectra.  Final article will only include active region, the other two will be available on-line only.
    \label{fig:example_spectra}}
\end{figure}

\section{APPLYING THE UNFOLDING METHOD TO AN AIA DATA SET \label{aiamodel}}

In the previous section, we determined the best parameters for the inversion using an MHD model.  Due to the temperature limitations of the model, we were unable to evaluate the impact of having broader range of temperatures in the solar emitting plasma would have on the inversion.  In this section, we apply the unfolding method to COSIE data simulated from a full-Sun emission map derived from AIA data that spans a larger range of temperatures.  We focus on how changing the range of temperatures of the inversion impacts the results. 

\subsection{A description of the data}

We prepared a full sun emission measure distribution as a function of position and over the range log $T = 4.5-7.5$ [EMD$(x,y,T)$]. We started with publicly available processed AIA observations (``Level-1 data'') of the Sun on 2015-Dec-28 around 13:15 UTC. To minimize the effect of uncertainties in the solar coronal relative abundances (see e.g. \citealt{odwyer_etal:2010,delzanna_etal:2011}), we restricted the dataset to the six EUV channels that are dominated by Fe lines (9.4\,nm, 13.1\,nm, 17.1\,nm, 19.3\,nm, 21.1\,nm, and 33.5\,nm).  To reduce the number of pixels to solve, and to increase the signal-to-noise ratios, we rebinned the data to a platescale of 4.8 arcseconds. The data were then deconvolved with the standard point spread functions
matched to the level of resolution.

For each pixel, we generated a set of EMDs of different temperature combinations, using the AIA effective areas distributed through SolarSoft together with a CHIANTI atomic model with a constant pressure of $10^{15}$ cm$^{-3}$ K$^{-1}$ and the \citet{Feldman92} coronal abundances. For many pixels, we were able to generate a set of EMDs that perfectly reconstructed the AIA data values. Where perfect solutions were not possible (due to errors in the measurements, instrument model, and/or atomic model), we generated sets of EMDs that reconstructed the AIA data to within $\chi^2 < 2.0$. To create a single representative EMD for each pixel, we eliminated members with higher emission measure weighted temperatures (i.e., $<T>_{\mathrm{EM}}$), and then took the mean EMD of the remaining members.    

At this point, we had an EMD map of the AIA observations, binned to a platescale of 4.8 arcseconds. This map was then centrally embedded into a larger array corresponding to the larger field of view of COSIE-C. The extra pixels were initially set to zero. However, we wished to extend the corona beyond the AIA field of view in order to test the sensitivity of COSIE. In addition, the performance of the unfolding might be affected by the overlapping effects of an extensive off-limb corona. To create an extended corona, the off-limb AIA corona between about 1.14 and 1.21 $R_\sun$ was partitioned into 14 annuli. This radial range was chosen to reduce the predominance of low-lying bright features, as well as avoiding the vignetted corners of the AIA field of view. A line fit was made to the radial dropoff in log(EMD). The corona beyond the AIA field of view was then filled in with random samples from the annular selected region, scaled to the radial dropoff function. Because the selected annular region did not completely exclude coherent bright structures, the sampling produced values that appear discontinuous with the diffuse corona at the edge of the AIA field of view, but the levels are properly understood as sampling both diffuse and bright, structured off-limb corona. Thereby, in some statistical sense, the artificial halo represents the extension of both the diffuse corona and bright structures (e.g., helmet streamers). For the particular analyses reported in this paper, in a final step the map was rescaled to the nominal COSIE-C platescale of 3.1 arcseconds.

Unfortunately, there remained unrealistic hot components in the EMD above Log T of 6.7 where the data are poorly constrained by the AIA data.   When calculating the COSIE intensities from the EMD over the entire temperature range, significant Ca XVII and Fe XXIV were present at levels that would be associated with a solar flare, instead of an active region (see expected values in Table~\ref{tab:linelist}).   To calculate the COSIE S+C data for this inversion, then, we use only the portion of the EMD maps below Log T = 6.7.  The COSIE S+C data derived from this data set is shown in Figure~\ref{fig:cosie_sc_truth_aia}.  Because there is no density information in the emission measure maps, the COSIE data was calculated assuming a constant pressure of $10^{15}$ cm$^{-3}$ K\none.  

\begin{figure}[ht]
    \centering
    \resizebox{.8\textwidth}{!}{\includegraphics{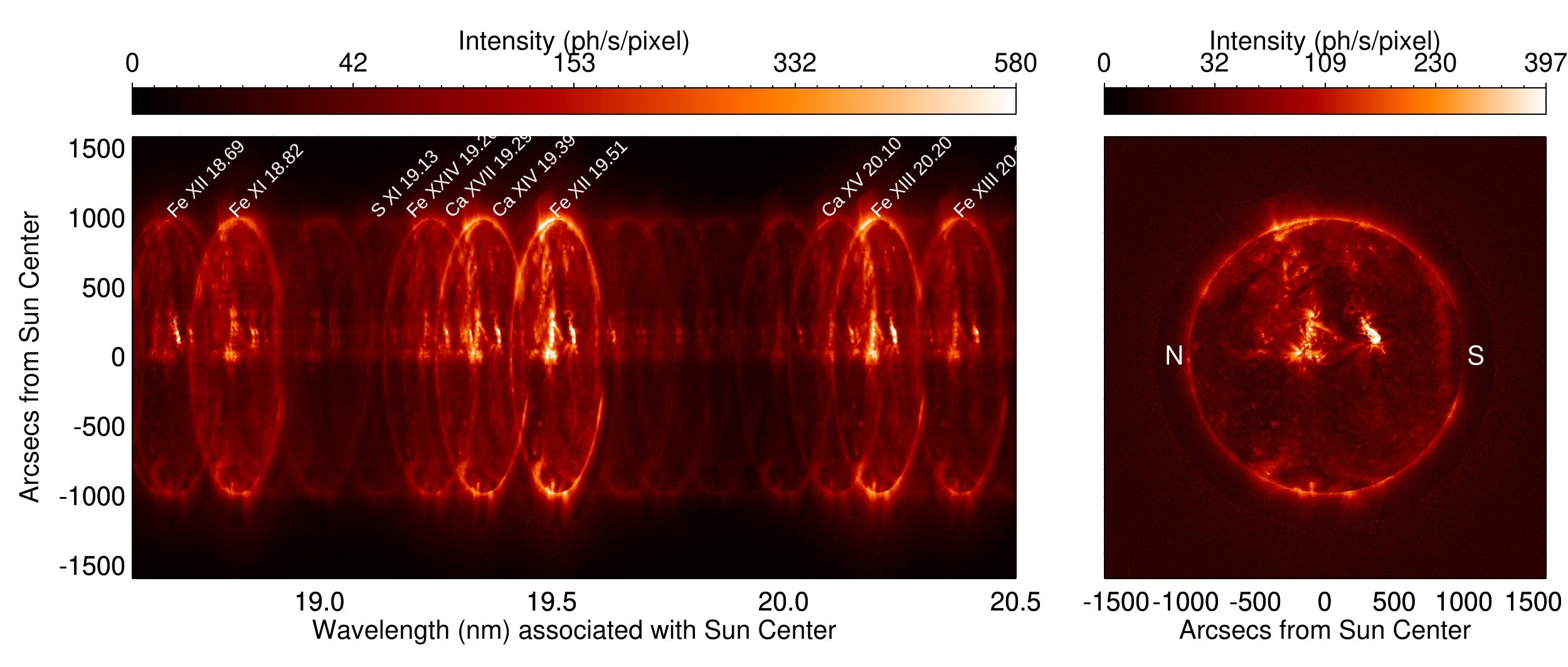}}
    \caption{The COSIE spectroheliogram (left) and coronagraph images associated with the emission measure distribution derived from the AIA data.  Only the central portion of the detector is shown.  The assumed exposure time is 1 s and noise has been added.  The Sun is oriented with North to the left of the coronagraph image.}
    \label{fig:cosie_sc_truth_aia}
\end{figure}

\subsection{Determining the best temperature range}

To unfold the AIA data, we use the line of sight and temperature resolution found in Section~\ref{mhdmodel}, namely  18.6\arc\  and 0.1.    We consider inversions at both constant pressure and constant density and ignore density in the inversion.   Details of the parameter space study are given in Table~\ref{tab:tempmap_aia}.  We first use the temperature range used in Section~\ref{mhdmodel}, 5.8$\leq$ Log T $\leq$ 6.4.  Those are runs A0-A3.  In these runs, the lowest $\chi^2$ is associated with a constant pressure of $10^{15}$ cm$^{-3}$ K\none or a constant density of $10^9$ cm$^{-3}$.  It is unsurprising  that both these solutions have low $\chi^2$ given that the data were calculated with a constant pressure assumption at $10^{15}$ cm$^{-3}$ K\none\ and most of the density sensitivity in the wavelength range is at Log T = 6.2, which would be consistent with a constant density of Log n = 8.8.

\begin{deluxetable}{ccccc}
\tablecaption{Summary of Runs for Temperature Maps from AIA data \label{tab:tempmap_aia}}
\tablehead{
\colhead{Run} & \colhead{Log T } & \colhead{Constant}  & \colhead{Constant} & \colhead{$\chi^2$}  \\
\colhead{Number}  & \colhead{Range}  & \colhead{Pressure}  & \colhead{Density} & \colhead{COSIE-S} \\
\colhead{} & \colhead{} & \colhead{(K cm$^{-3}$)} &   \colhead{(cm$^{-3}$)}& \colhead{}   
}
\startdata
A0   &   5.8-6.4  &  15  &  N/A  &  4.1  \\  
A1   &   5.8-6.4  &  16  &  N/A  &  8.3  \\  
A2   &   5.8-6.4  &  N/A  &  8  &  8.3  \\  
A3   &   5.8-6.4  &  N/A  &  9  &  4.3  \\  
A4   &   5.3-6.4  &  15  &  N/A  &  8.6  \\  
A5   &   5.6-6.4  &  15  &  N/A  &  6.1  \\  
A6   &   5.6-6.6  &  15  &  N/A  &  9.9  \\  
A7   &   5.6-6.8  &  15  &  N/A  &  18.0  \\  

\enddata
\tablecomments{All calculations are made for a spatial range of $\pm$ 1.2 solar radii, a LOS resolution of 18.6\arc, and a temperature resolution of $\Delta$ Log T = 0.1.}
\end{deluxetable}

Next, we run the inversion over a larger temperature range for the constant pressure of $10^{15}$ cm$^{-3}$ K\none.  The parameters used in the inversion and the $\chi^2$ of the solutions are given in Table~\ref{tab:tempmap_aia} as Runs A4-A7, with the temperature ranges considered in the second column.  First we consider extending the temperature range to lower temperatures (Run A4 and A5).  Recall that the COSIE wavelength range includes an O V and O VI line formed at Log T 5.35 and 5.45, respectively (see Table~\ref{tab:linelist}), so we first expand the temperature range down to Log T = 5.3 (Run A4). However the count rates in the O lines are expected to be weak except in a solar flare and there are no additional diagnostics between the O lines and the Fe VIII line formed at Log T = 5.65.  In Run A5, we expand the minimum temperature range to Log T = 5.6, which would include the temperature of peak emissivity for the stronger Fe VIII spectral line.  Next we increase the maximum temperature considered in the inversion.  Again, there are few spectral lines in the COSIE wavelength range that can constrain the emission measure distribution above 6.3.  The Ar XIV and Ca XIV, XV and XVII lines are expected to be weak.  In Runs A6-A7, we increase the maximum temperature to Log T = 6.6 and 6.8, respectively.

Our study finds that increasing the temperature range does not improve the normalized $\chi^2$.  This could be due to two reasons.  Because there are more free parameters, the normalization factor for the $\chi^2$ gets smaller.  Also there could be rows that do not converge to acceptable solutions.  Regardless, it does not appear that increasing the temperature range improves the agreement with the data.  We proceed, then, with the same temperature range used in Section 4.   

Figure~\ref{fig:cosie_sc_inv_aia} shows the inverted spectrometer and coronagraph images (top panels) and the difference between the inverted and true data shown in Figure~\ref{fig:cosie_sc_truth_aia} in the bottom panels for Run A0 (temperature range of 5.8$\leq$ Log T $\leq$ 6.4).  Note that the truth data includes emission measure at higher and lower temperatures than considered in the inversion.  Spectral lines emitted at those higher or lower temperatures would not be present in the inverted data.  Finally, we compare the truth and inverted emission measure maps in Figure~\ref{fig:tempmap_aia} and Fe XII and XIII intensities in Figure~\ref{fig:comp_moments_fe12_aia}.

\begin{figure}[ht]
    \centering
   \resizebox{.8\textwidth}{!}{\includegraphics{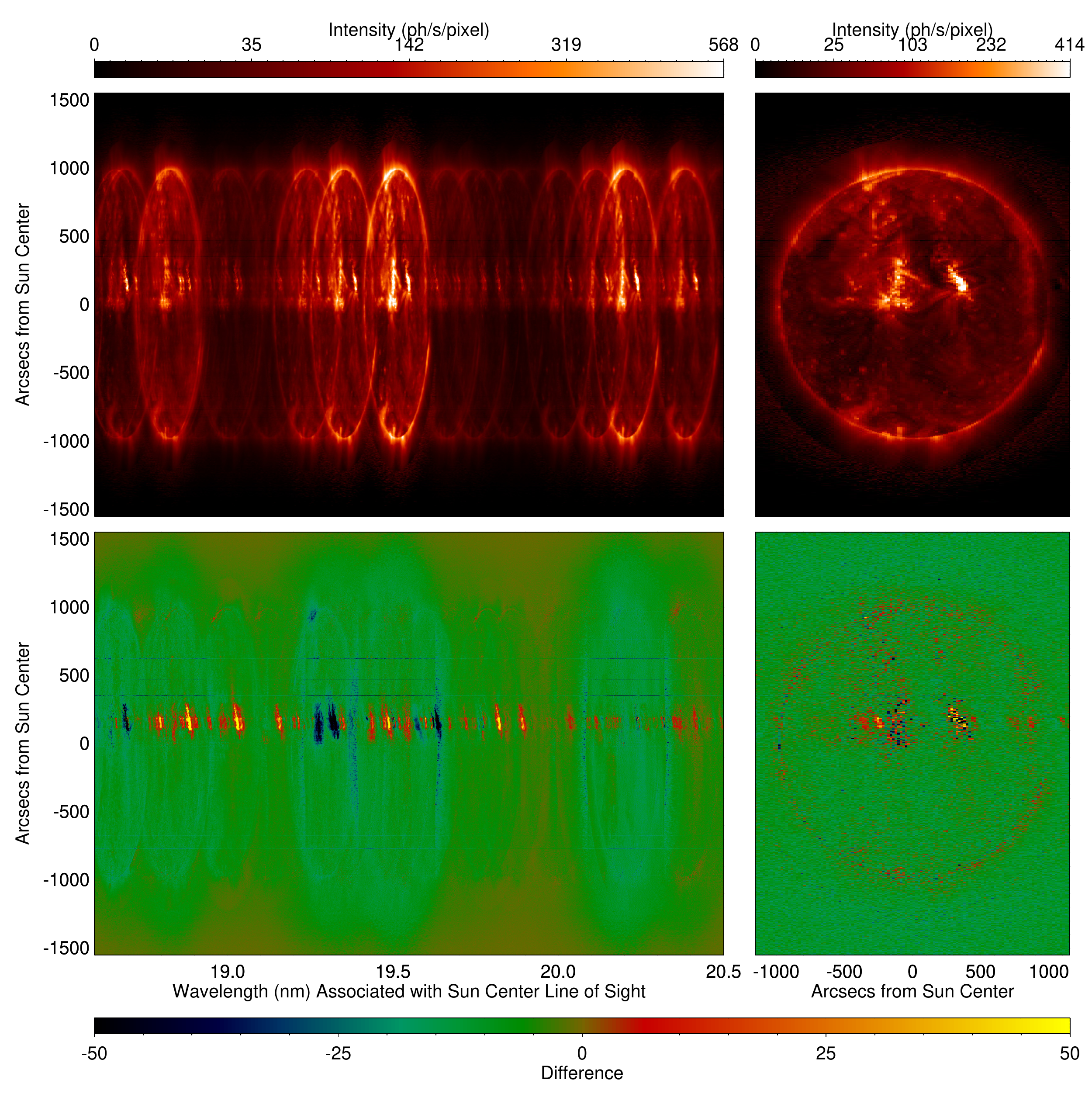}}
    \caption{The COSIE spectroheliogram (left) and coronagraph images from Run A0.  The bottom panels show the difference in the inverted and truth data shown in Figure \ref{fig:cosie_sc_truth_aia}.}
    \label{fig:cosie_sc_inv_aia}
\end{figure}

\begin{figure}[ht]
    \centering
    \resizebox{.8\textwidth}{!}{\includegraphics{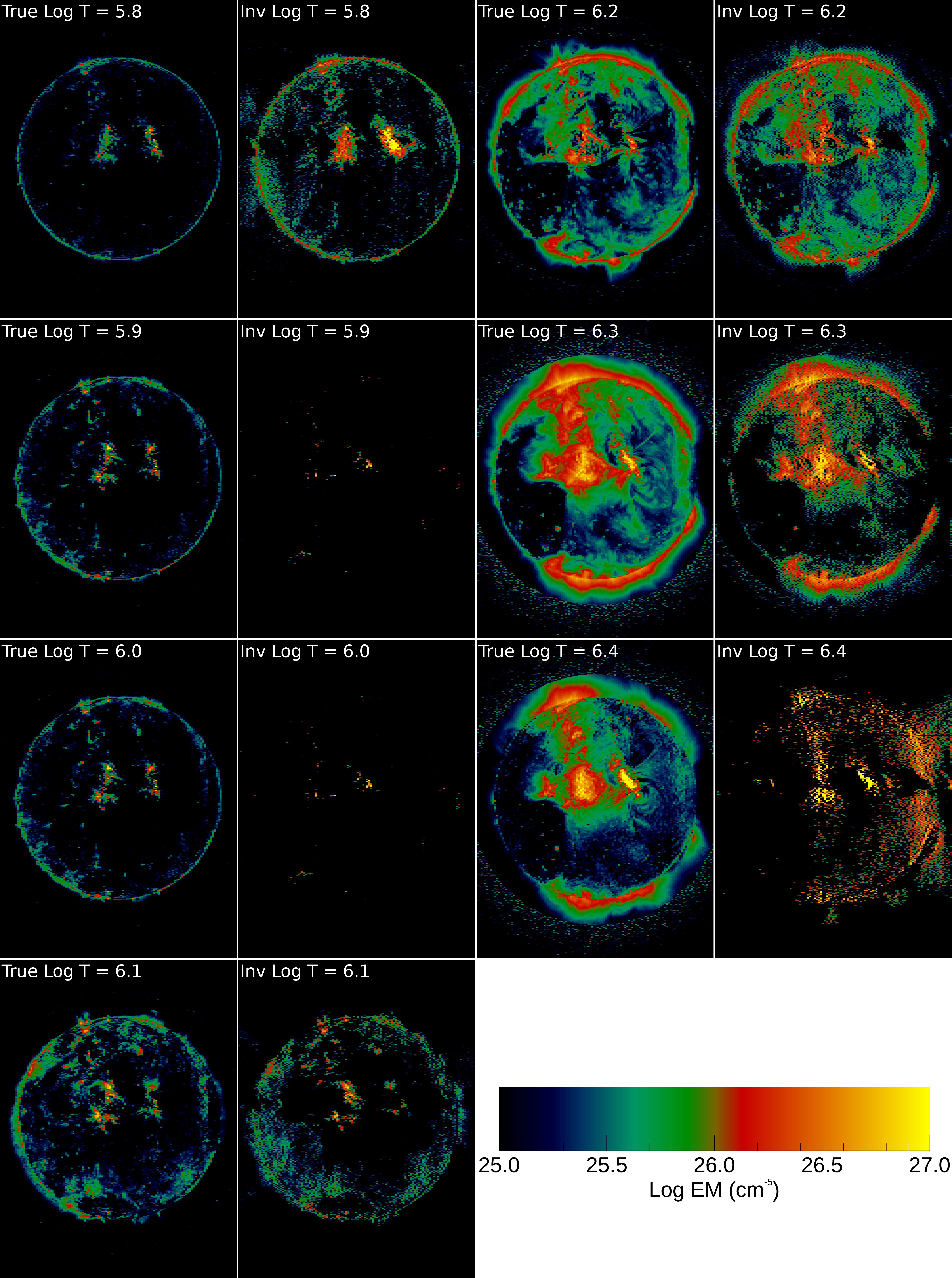}}
    \caption{The true emission measure distribution in different Log T bins 5.8-6.4 is shown in the first and third columns, the inverted EM distribution in the same temperature bins is shown in the second and fourth columns.}
    \label{fig:tempmap_aia}
\end{figure}

\begin{figure}[ht]
    \centering
    \resizebox{.8\textwidth}{!}{\includegraphics{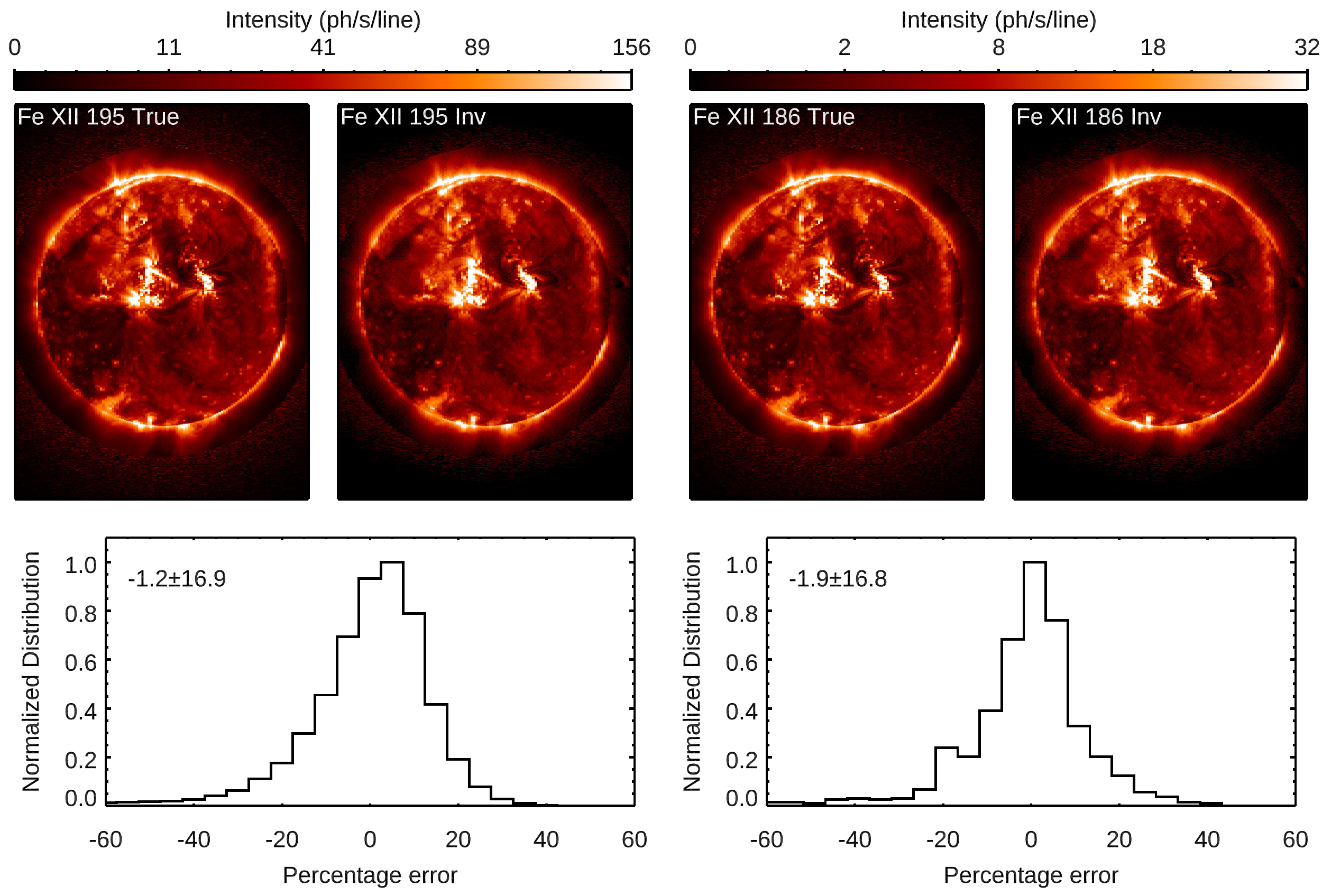}}
     \resizebox{.8\textwidth}{!}{\includegraphics{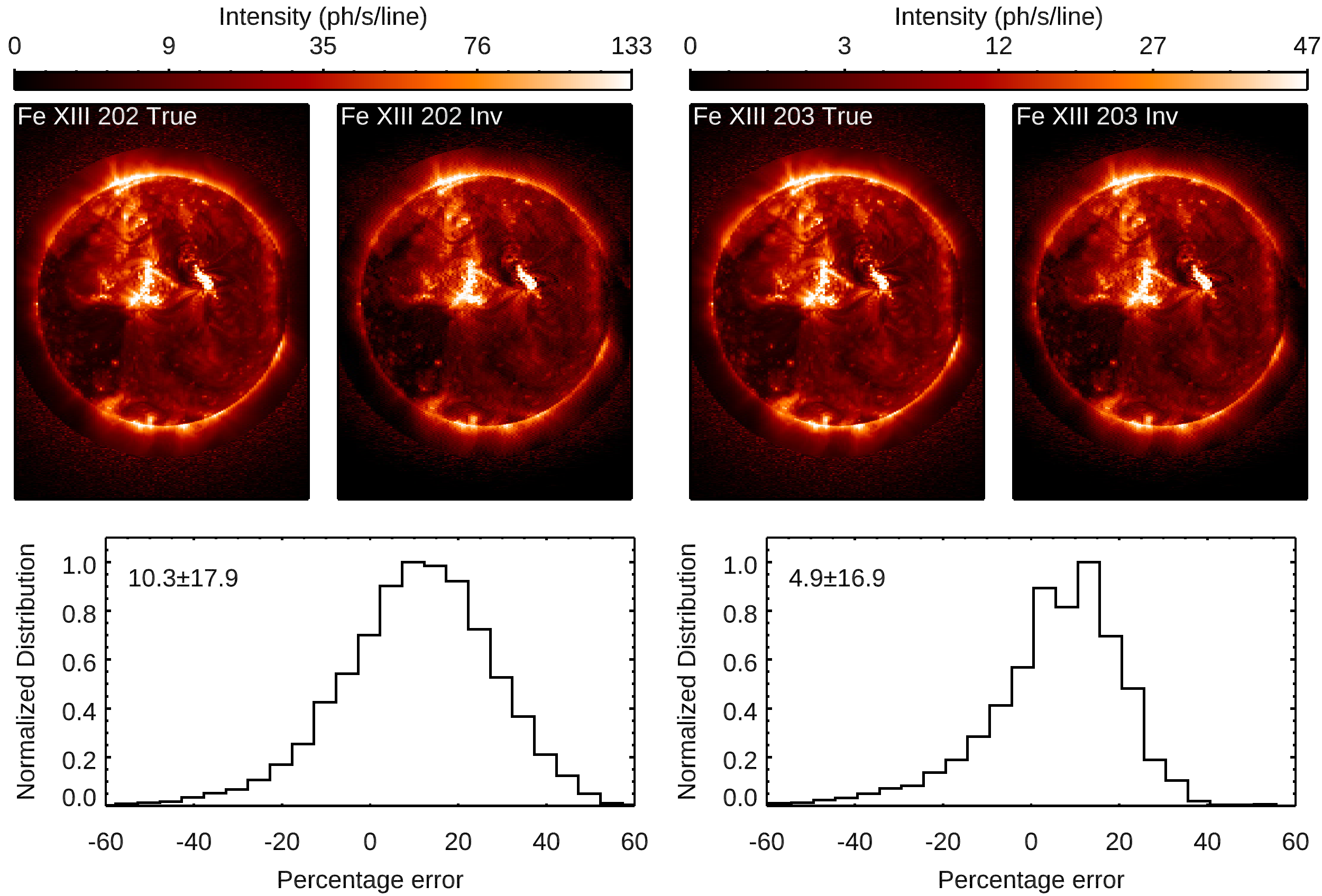}}
    \caption{A comparison of the true and predicted Fe XII (top) and Fe XIII (bottom) lines.  Intensity maps are scaled to the 0.5 power.  The histograms reflect the percentage errors for all pixels with greater than 30 photons s\none\ line\none.  The average and standard deviation of the percentage error is given in each line plot. 
    \label{fig:comp_moments_fe12_aia}}
\end{figure}


\section{DISCUSSION AND CONCLUSIONS \label{sec:disc}}

In this paper, we have demonstrated an unfolding technique for data acquired with a slitless spectrometer and imager.  We have used simulated data from the proposed COSIE instrument to optimize this technique for emission measure maps at constant temperature, maps of spectrally pure intensity in the Fe XII and Fe XIII lines and density maps based on both Fe XII and Fe XIII diagnostics.  For this instrument, we found that we could invert the data with a spatial resolution in the dispersive direction of 18.6\arc\ (two times worse than the nominal spatial resolution of the spectrometer), while maintaining the 3.1\arc\ resolution in the cross-dispersive direction.  The optimal temperature resolution was found to be $\Delta$ Log T = 0.1 over a temperature range of 5.8 $\leq$ Log T $\leq$ 6.4 for a quiescent, non-flaring Sun.  These parameters are closely coupled with the instrument design. One can imagine a different design that would have better or worse spatial resolution, and, if using different EUV passbands, would have sensitivity to different temperature ranges. 

The forward and inversion modelling we have used for the MHD model and the AIA dataset has been very useful to test the methods but is far from the real case. To further test and refine the modelling, we plan to work on real spectra obtained  from Hinode EIS. Regarding the present models, they are largely independent of the quality of the underlying atomic data. The reliability of the present modelling for real COSIE-S data will instead of course depend not just on the accuracy but also on the completeness of the atomic data. Over the past 10 years, a series of detailed papers (see the references in \citealt{delzanna_mason:2018}) have shown new atomic data calculations and new spectral lines identifications.  These data have been benchmarked against the high-resolution Hinode EIS spectra, which observed the COSIE-S wavelength range.   This implies that for the COSIE wavelength range, the atomic data, as made available with CHIANTI version 8 \citep{delzanna_etal:2015}, are fairly complete and very accurate. In particular, the Fe XII and Fe XIII density diagnostics available to COSIE-S are excellent ones.  We are therefore confident that the present modelling will be able to provide reliable information on temperatures and densities across the whole Sun, on-disk and off-limb with unprecedented temporal cadence. The benchmark of the atomic data against the high-quality Hinode EIS spectra has also indicated that several transitions, even at such high resolution, are blended in very different ways, depending on what is observed.  However, in most cases there is sufficient information in these wavelengths to deblend the lines. 

As all the strongest COSIE-S lines for any  solar conditions (e.g. quiet Sun, coronal holes, active regions, bright points, flares) are from Fe, the present modelling (together with the derived temperatures and densities) is largely independent of the choice of chemical abundances. However, we note that the Fe abundance does vary from solar region to region, so derived quantities such as emission measures should be treated with caution within any science analysis. As reviewed in \cite{delzanna_mason:2018}, there is now considerable evidence that in the cores of active regions the Fe abundance is about a factor of 3.2 higher than its photospheric value. On the other hand, active region loops of about 1 MK have a range of values, while Fe has nearly photospheric abundances in coronal  holes and most likely also in the quiet solar corona.  

In this paper, we have focused on a quiescent, non-flaring Sun.  COSIE is not sensitive to the magnitude of the velocities in the MHD simulation ($<$ 50 km s$^{-1}$) and there is of course no velocity information in the AIA dataset.  In our next paper, we will use an MHD simulation of a CME eruption with much larger velocities to investigate the sensitivity of COSIE in the velocity parameter space. 

One of the key aspects of the success of this inversion method is that we are able to couple the spectrometer data with the spectrally-integrated image data.  The image data is able to constrain the spatial distribution of the emission, while the spectrometer data constrained the temperature and density distribution of the emission.  Without the image data, we would have been far less successful. 

Finally, we plan on applying this technique to other modern data sets that include data from slitless spectrometers, such as MOSES \citep{Kankelborg01,fox10}. Applying it to other data sets implies defining $\mathbold{M}$ for that instrument and performing a similar parameter space and sensitivity study that is included in this paper for COSIE.  For instance, MOSES is tuned to observe the He II 30.4 nm line, with contributions from a few higher temperature lines in the passband. In that case, we would expect to not be able to constrain the emission measure as a function of temperature, but instead focus on the velocity distribution of the emission measure.

\appendix

\section{Defining $\mathbold{M}$ for COSIE \label{definem}}

$\mathbold{M}$ is the matrix that maps the emission measure ($EM = n^2dz$) along a specific line-of-sight (LOS) to the Sun into detector pixels.  In the COSIE spectroheliogram, emission measure along a  single LOS maps into multiple pixels across the detector, while for the coronagraph, emission measure along a single LOS maps into a single  pixel.  In this section we describe how we define and calculate $\mathbold{M}$.  

The COSIE detector has 2k detector pixels, $n_{det}$, across a row, so  $\mathbold{y_{obs}}$ is then a $[2*n_{det}]$ vector where the first $n_{det}$ is the spectrometer data and the second $n_{det}$ is the coronagraph data.  The emission measure matrix, $\mathbold{x}$, must be defined at whatever temperature, density, velocity, and spatial locations that contribute to the spectrometer and coronagraph data; we currently call the total number of free parameters in emission measure $n_{em}$ and revisit shortly how these values are determined.  Finally, the matrix $\mathbold{M}$ is $[n_{em}\times 2*n_{det}]$, where $M{ij}$ describes how the emission measure with a specific line of sight, temperature, density and velocity, $EM_j$, maps into detector pixel $i$.

Each row of $\mathbold{M}$ is the thermally-broadened isothermal, isodense spectrum generated with spectral line emissivities from CHIANTI version 8 \cite{delzanna_etal:2015} and folded with the COSIE-S+C effective areas and mapped to the appropriate detector positions for line of sight location.   Figure~\ref{fig:cosie_example} shows rows of $\mathbold{M}$ for several different conditions.   The top panel shows how an emission measure of $10^{27}$ cm$^{-5}$ from a LOS to sun center with Log T = 6.2 and log n = 9.0 maps into the detector for the spectrometer (left) and coronagraph (right).  The expected intensity as a function of detector position is given in units of photons s$^{-1}$ pixel$^{-1}$.    For the spectrometer, the wavelength array on the x-axis corresponds to the Sun center location.  For the spectrometer, the response is a spectrum that contains several spectral lines, the strongest line is the Fe XII line is at 19.5 nm.  For the coronagraph, the signal is confined to a single coronagraph pixel.  These two responses together (S+C) constitutes one row of $\mathbold{M}$.  Another row of $M$ might describe how $EM$ from a LOS along the east limb with the same temperature and density maps into the detector, this is shown in the middle panels of Figure~\ref{fig:cosie_example}.  It is essentially the same spectrum as the top panel shifted to the left. Another row of $\mathbold{M}$ is how $EM$ from Sun center with Log T = 5.8 and Log n = 8.0 maps into the detector, this is shown at the bottom panels of Figure~\ref{fig:cosie_example}.  Since this spectrum is at a different temperature, a different set of spectral lines is predicted. 

\begin{figure*}[ht]
    \begin{center}
    \resizebox{.49\textwidth}{!}{\includegraphics{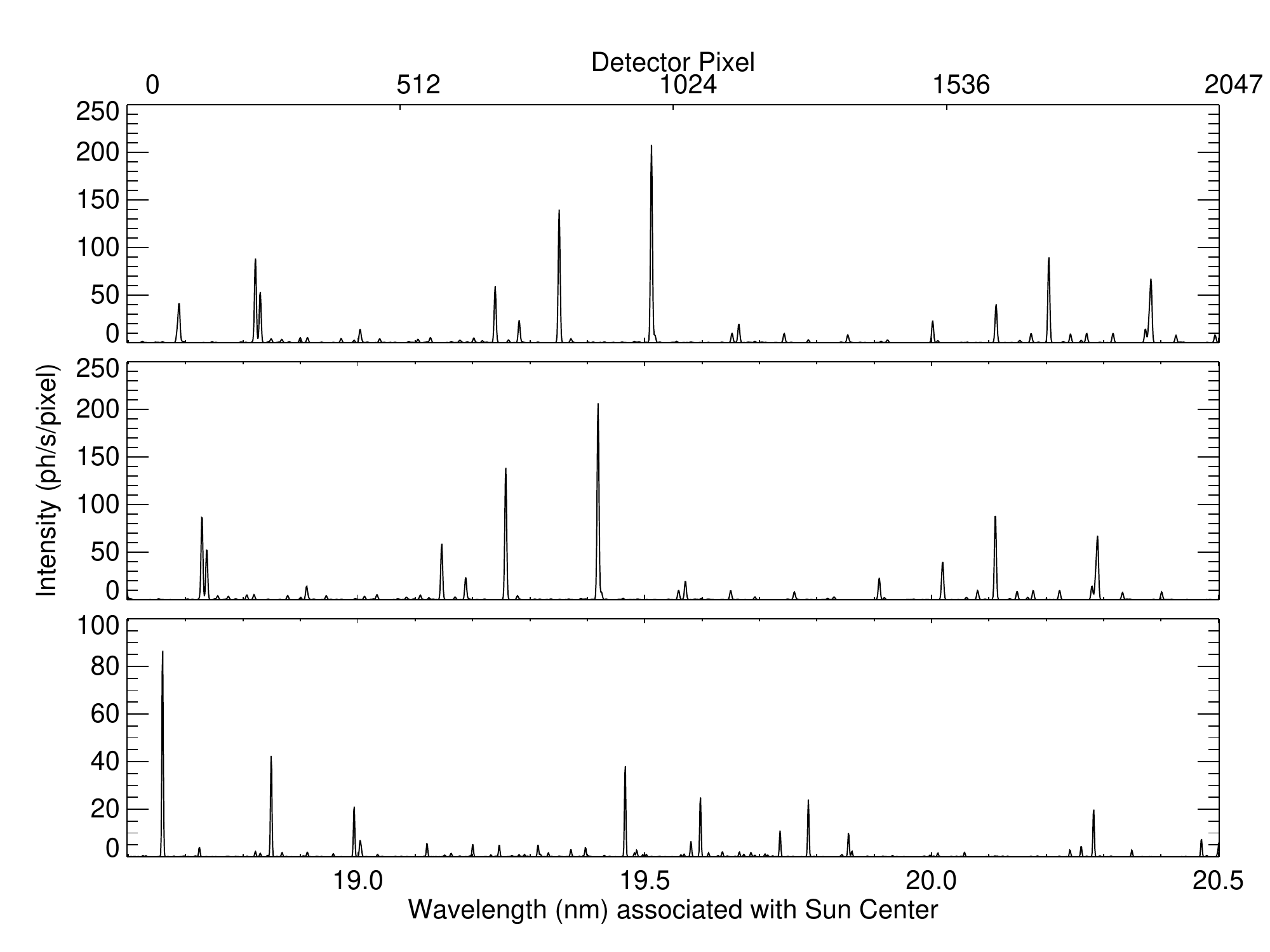}}
    \resizebox{.49\textwidth}{!}{\includegraphics{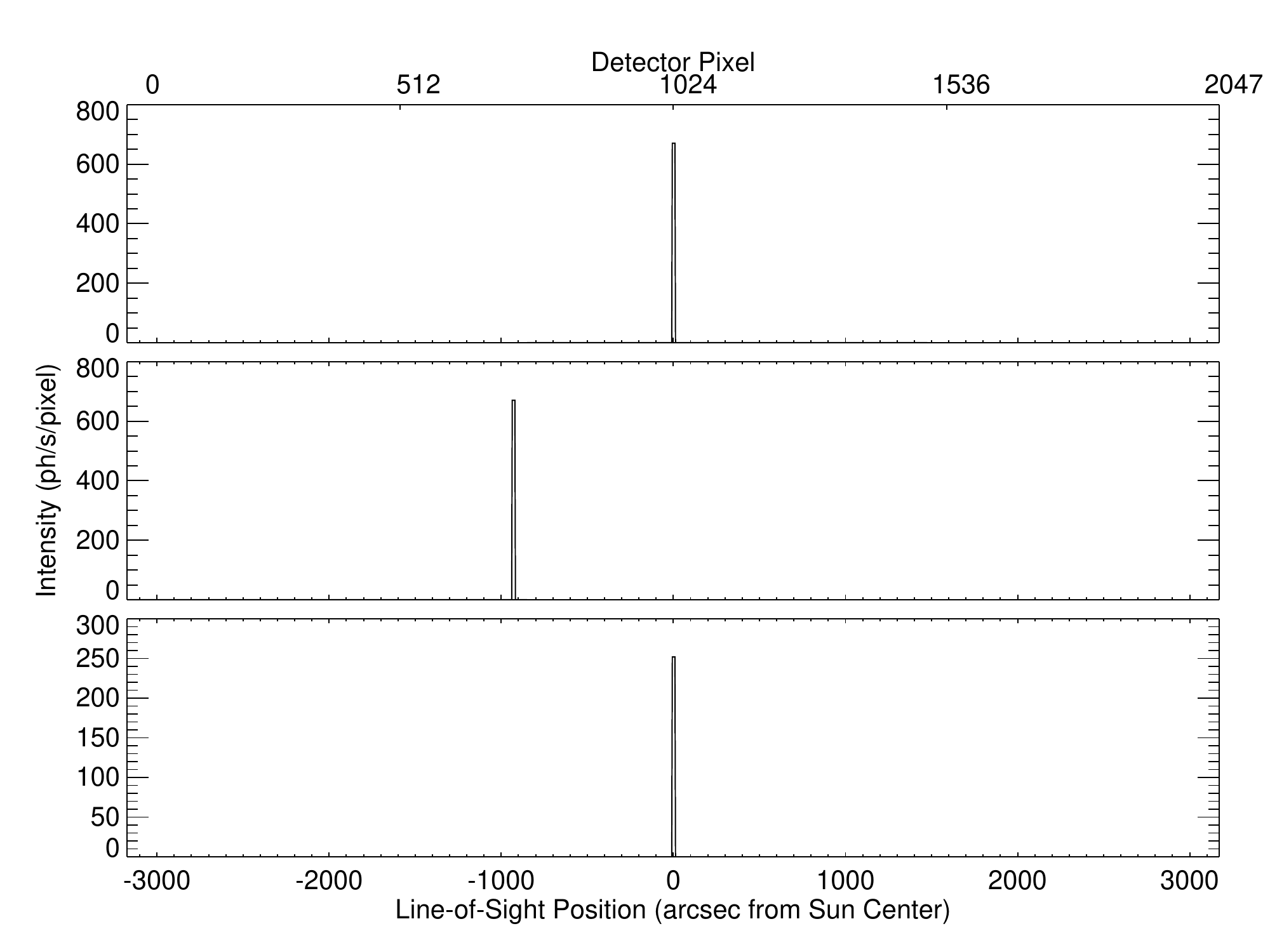}}
    \caption{The COSIE spectrometer (left) and coronagaph (right) responses for $EM = 10^{27}$ cm$^{-5}$ at three different values of LOS, Log T and Log n.  Top panels: A LOS along Sun center with Log T = 6.2 and Log n = 9.0.  Middle panels: A LOS along the east limb of the Sun with Log T = 6.2 and Log n = 9.0.  Bottom panels:  A LOS along Sun center with Log T = 5.8 and Log n = 8.0.}
    \label{fig:cosie_example}
    \end{center}
\end{figure*}

\acknowledgments

This work was supported at SAO and MSFC by NASA Grant 80NSSC18K0197 for COSIE Technology Development.
CD acknowledges research support from NASA (HSR and LWS programs) and AFOSR. Computational support provided by NASA’s Advanced Supercomputing Division, NSF’s Texas Advanced Computing Center, and San Diego Supercomputer Center.
GDZ acknowledges support from STFC (UK) via the consolidated grant 
to the atomic astrophysics group at DAMTP, University of Cambridge and from SAO during his visit to CfA. 
This material is based upon work supported by the National Science Foundation under Grant No. 1460767 that supported AA and CC during the University of Alabama-Huntsville and MSFC  Research Experiences for Undergraduate program.

\bibliographystyle{aasjournal}

\bibliography{unfold,bib_extra_CD}

\end{document}